\documentclass[fleqn,usenatbib]{mnras}

\usepackage{mathptmx}
\usepackage{natbib}
\usepackage[T1]{fontenc}

\DeclareRobustCommand{\VAN}[3]{#2}
\let\VANthebibliography\thebibliography
\def\thebibliography{\DeclareRobustCommand{\VAN}[3]{##3}\VANthebibliography}

\usepackage{graphicx}	
\usepackage{amsmath}	
\usepackage{amssymb}	
\usepackage{xcolor}
\usepackage{ulem}

\title[Giant flares at high latitudes]{Giant white-light flares on fully convective stars occur at high latitudes}

\author[Ekaterina Ilin et al.]{Ekaterina Ilin$^{1,2}$\thanks{E-mail: eilin@aip.de (EI)}, Katja Poppenhaeger$^{1,2}$, Sarah J. Schmidt$^{1}$, Silva P. J\"arvinen$^{1}$,\newauthor Elisabeth R. Newton$^{3}$, Juli\'an D. Alvarado-G\'omez$^{1}$, J. Sebastian Pineda$^{4}$,  \newauthor James R. A. Davenport$^{5}$, Mahmoudreza Oshagh$^{6,7}$, \& Ilya Ilyin$^{1}$
\\
\\
$^{1}$Leibniz Institute for Astrophysics Potsdam (AIP), An der Sternwarte 16, 14482 Potsdam, Germany\\
$^{2}$Institute for Physics and Astronomy, University of Potsdam, Karl-Liebknecht-Str. 24/25, 14476 Potsdam, Germany\\
$^{3}$ Department of Physics and Astronomy, Dartmouth College, Hanover, NH 03755, USA\\
$^{4}$ University of Colorado Boulder, Laboratory for Atmospheric and Space Physics, 3665 Discovery Drive, Boulder CO, 80303, USA \\
$^{5}$ Department of Astronomy, University of Washington, Seattle, WA 98195, USA\\
$^{6}$ Instituto de Astrof\'isica de Canarias (IAC), 38205 La Laguna, Tenerife, Spain \\
$^{7}$ Departamento de Astrof\'isica, Universidad de La Laguna (ULL), 38206, La Laguna, Tenerife, Spain
}

\date{Accepted 2021 July 19. Received 2021 July 7; in original form 2021 February 24}

\pubyear{2021}

\begin{document}

\defcitealias{davenport2014}{D14}
\label{firstpage}
\pagerange{\pageref{firstpage}--\pageref{lastpage}}
\maketitle

\begin{abstract}
White-light flares are magnetically driven localized brightenings on the surfaces of stars. Their temporal, spectral, and statistical properties present a treasury of physical information about stellar magnetic fields.
The spatial distributions of magnetic spots and associated flaring regions help constrain dynamo theories. Moreover, flares are thought to crucially affect the habitability of exoplanets that orbit these stars. Measuring the location of flares on stars other than the Sun is challenging due to the lack of spatial resolution.
Here we present four fully convective stars observed with the Transiting Exoplanet Survey Satellite (TESS) that displayed large, long-duration flares in white-light which were modulated in brightness by the stars' fast rotation. This allowed us to determine the loci of these flares directly from the light curves. All four flares occurred at latitudes between $55^\circ$ and $81^\circ$, far higher than typical solar flare latitudes.
Our findings are evidence that strong magnetic fields tend to emerge close to the stellar rotational poles for fully convective stars, and suggest that the impact of flares on the habitability of exoplanets around small stars could be weaker than previously thought.
\end{abstract}




\section{Introduction}
\label{sec:intro}
Cool stars, i.e.\ stars with an outer convective envelope, are known to display a variety of magnetically driven phenomena, such as starspots, flares, and coronal mass ejections, collectively called magnetic activity~\citep{schrijver2000}. Sun-like stars possess a radiative core, and their magnetic activity is understood to be driven by a magnetic dynamo of the $\alpha$-$\Omega$ type~\citep{parker1955,roberts1972}. For low-mass stars that are fully convective~(spectral type M4V and later,~\citealt{stassun2011}), a type of dynamo must be operating that does not require the presence of a radiative core. Surprisingly, the relation between rotation and magnetic activity for stars on either side of the fully-convective boundary is the same in both X-ray~\citep{wright2016,wright2018} and H$\alpha$ \citep{newton2017} observations, suggesting that the dynamo operating in fully convective stars has a similar efficiency as the one in our Sun.
One important difference between Sun-like stars and low-mass M dwarfs is that Sun-like stars spin down quickly, while M dwarfs remain fast rotators much longer~\citep{irwin2011,newton2016}. Fast rotation fuels enhanced magnetic activity and therefore frequent flaring, which is exploited in studies of stellar flares in M dwarf stars down to the bottom of the main sequence~\citep{stelzer2006,robrade2010,schmidt2016,paudel2018}.

On the Sun, emerging magnetic fields drive flares, and stronger fields drive more energetic flares, making them highly localized markers of dynamic magnetic field concentrations~\citep{priest2002}.
In optical light curves, flares manifest as sudden brightenings that last between a few minutes and several hours. The onset of a flare is characterized by a fast rise in stellar flux which peaks in a sharp maximum, followed by an exponential, and then a more gradual, decay back toward the quiescent state.

The location of flares on the stellar surface constrains the surface field strength and topology, and tests predictions from stellar dynamo theories. On the Sun, the positions of large flares follow the cyclic latitude variations of sunspots~\citep{zhang2007}. The latter is known as the butterfly diagram, and is crucial to the understanding of the solar dynamo~\citep{gnevyshev1977}. Fast-rotating young solar-like stars and low mass stars have been predicted to form magnetic spots close to the rotational poles~\citep{schuessler1992,granzer2000, gastine2013, yadav2015}, and this prediction has been confirmed by spectroscopic and spectropolarimetric observations~\citep[e.g.,][]{strassmeier1998, marsden2006, jarvinen2007, jarvinen2008, isik2018, barnes2015, barnes2017}. If these spots are co-spatial with large flares, they may be classified as intensely active regions with a dynamic local field configuration.

Differences in flaring region loci can also, at least partially, explain the variety seen in flare light curve morphologies. The convex shape of a flare light curve in X-rays on V773 Tau~\citep{skinner1997} could be well-described by rotational modulation. Later, effects of rotation and eclipses have been put forward as an explanation for unusual flare light curves multiple times~\citep{stelzer1999,montmerle2000,johnstone2012}. However, only few examples of unambiguously located stellar flares exist to date~\citep{schmitt1999,wolter2008, peterson2010}. 

Here, we present observations of four giant flares that occurred on fast-rotating fully convective dwarfs detected in photometry from the Transiting Exoplanet Survey Satellite (TESS,~\citealt{ricker2015}). These flares are remarkable because they lasted longer than a full rotation period of the star so that the characteristic flare light curve was modulated by the flaring footpoint's movement in and out of view on the stellar surface. In Section \ref{sec:data}, we describe our selection of both the stellar and the flare sample, and explain how we determined stellar properties for the promising candidates. In Section \ref{sec:methods}, we introduce a model of large flares with rotationally modulated light curves, show how we can use it to infer the latitude of the flaring regions, and discuss the assumptions made in the model. We fit our model to the observations, and present the inferred flaring region latitudes in Section \ref{sec:results}. We briefly discuss quasiperiodic pulsations as an alternative model, place our results in the context of current research on stellar surface magnetic fields and dynamo thoeries for fully convective stars, and explore their implications for exoplanet habitability in Section \ref{sec:discussion}. We conclude with a short summary and an outlook in Section \ref{sec:summary}.

\section{Data}
\label{sec:data}We started our systematic search for long-duration superflares on fully convective stars with the entire TESS archive. First, we used a color and brightness cut to restrict our search to fully convective nearby stars (Section \ref{sec:data:stars}). Second, we searched their 2 minute cadence light curves for flares, and selected those candidate events where we could detect a rotation period that was shorter than the visible flare decay (Section \ref{sec:data:flares}). Our final sample consists of four giant flares on four stars with spectral types M5-M7 and rotation periods between 2.7 and 8.4 hours. Two stars in our sample lacked measurements of projected rotation velocity in the literature. For those we obtained high-resolution spectra~(Section \ref{sec:data:hires}), so that we could determine the properties of all stars, including the inclination of the stellar rotation axis~(Section \ref{sec:data:properties}).
\subsection{Stellar sample and TESS light curves}
\label{sec:data:stars}
We compiled a sample of nearby fully convective dwarf stars of spectral type M5 and later that were observed with TESS in the first two years of observations (Cycles 1 and 2). We selected bright stars with TESS magnitude $<15$, which we crossmatched within 5 arcsec with nearby stars (parallax $> 20$ arcsec) in Gaia~\citep{gaiacollaboration2016} DR2~\citep{gaiacollaboration2018}. We required good $G$ and $RP$ colour measurements, i.e., band flux divided by its error (\texttt{flux\_over\_error} in the Gaia catalog) $>10$, and applied a color cut to select for fully convective stars. We applied $1.27<G-RP<1.75$, based on the relation between $G-RP$ band color and spectral type~\citep{kiman2019} derived from spectroscopically identified M and L dwarfs in the Sloan Digital Sky Survey~\citep{york2000} (SDSS). In total, we analysed 3,175 targets that were observed with TESS.

\subsection{Flare sample}
\label{sec:data:flares}
We searched for stars that displayed both long duration flares in their light curves, and clear rotational modulation outside of flares with a period shorter than the visible flare decay. Modulation of the flare itself was not a selection criterion. The typical TESS light curve spans one Sector, or 27 days, and we used only stars with 2-minute cadence data to ensure that we could fit the shapes of the superflares in sufficient detail. Our sample covers TESS Cycles 1 and 2, Sectors 1-26, encompassing both hemispheres above and below the ecliptic, that is, about $73$ per cent of the whole sky \citep{ricker2015}. We used the flare finding algorithm \texttt{AltaiPony}~\citep{altaipony2020} to detect flares on 6,291 light curves that were obtained for the 3,175 sample stars.

From the 17,646 automated flare detections we selected those with relative amplitudes above $20$ per cent. For each light curve with at least one such flare we derived the Lomb-Scargle periodogram~\citep{lomb1976, scargle1982} using \texttt{lightkurve}~\citep{lightkurve2018} in order to determine the stellar rotation period as the signal with the highest peak and $SNR>2.5$. For the promising cases, we compared the periodograms derived from the TESS light curves that were systematic corrected with the Presearch Data Conditioning pipeline~\citep[\texttt{PDCSAP\_FLUX},][]{jenkins2016}, and those extracted from the raw flux (\texttt{SAP\_FLUX}). We selected candidates where the measured flare duration was at least $15$ per cent of the detected rotation period. The duration was chosen to ensure that we did not exclude the superflares that were split into multiple detections by~\texttt{AltaiPony} due to rotational modulation. We arrived at this threshold by injecting model superflare events into simulated light curves with typical Gaussian noise levels, and recovering their properties using \texttt{AltaiPony}~(see also Section~\ref{sec:discussion:detection}).

\subsubsection{Manual inspection of flare events}
We identified 343 candidate events for superflares which we inspected visually. Out of those, in 19 cases the detected rotation period was a fluke in the periodogram where multiple peaks could not be separated into rotational and non-rotational signal. Some candidates were instrumental false positives (19), others were contaminations from Solar system objects that mimicked flare events (6), both of which could be disambiguated from the Target Pixel Files from which the light curves were extracted. The majority, however, were flares with total duration below one rotational period (291), partly because aliases in the periodogram suggested faster rotation, but mostly because the conservative $15$ per cent relative flare duration threshold allows selection of short flares. Finally, four flares lasted for multiple rotation periods but exhibited a complex flare morphology that indicated the presence of more than two superimposed flares in each event~(Table~\ref{tab:excl}).

We observed four long duration superflares, i.e., flares with TESS band energies above $10^{33}$ erg, on fully convective M dwarf stars in the red-optical regime (``white-light'') with TESS: TIC~277539431 (2MASS~J10551532-7356091), TIC~452922110 (2MASS~J08055713+0417035), TIC~44984200 (2MASS~J08380224-5855583), and TIC~237880881 (2MASS~J01180670-6258591), hereafter referred to as  TIC~277, TIC~452, TIC~449, and TIC~237.

\begin{table}
\centering
\caption{Long-duration flares on small stars with complex shapes that indicate the presence of more than two superimposed events. These targets were excluded from the analysis.}
\begin{tabular}{l|ccc}
\hline\hline
       TIC &  Sector &  $P$ [h] &  $t_{\text{start}}$ [BTJD - 2457000 d] \\
\hline
 441811894 &      18 &     5.13 &                            1804.987685 \\
 441606549 &      20 &     4.49 &                            1846.901948 \\
 298160985 &      14 &     6.90 &                            1704.145163 \\
 260972843 &       5 &     6.83 &                            1454.968234 \\
\hline

\end{tabular}

\label{tab:excl}
\end{table}

\subsection{High-resolution spectra}
\label{sec:data:hires}
We collected high-resolution spectra for two of our sample stars that did not have literature values for their projected rotation velocity $v\sin i$ available. TIC 449 and TIC 277 were observed with the High-Resolution Spectrograph~\citep[HRS,][]{crause2014} at the Southern African Large Telescope~\citep[SALT,][]{buckley2006} on February 08 and February 09 2020, respectively. The HRS provides spectra covering a wavelength range of $5,500-8,900\,$\AA$\;$ in its red arm with a spectral resolution of about $40,000$ in our chosen medium-resolution mode. Since our targets have low effective temperatures, the bulk of the stellar emission was observed redwards of $7,000\,$\AA$\;$. The data were reduced with the PEPSI data reduction software~\citep{2018Strassmeier}. The reduction followed the standard steps of bias overscan, detection and subtraction, scattered light extraction from the inter-order space and subtraction, definition of \'echelle orders, optimal extraction of spectral orders, wave-length calibration, and a self-consistent continuum fit to the full two-dimensional (2D) image of extracted orders. The one-dimensional spectra resulting from combining the \'echelle orders were used to infer the rotational line broadening as described in Section~\ref{sec:vsini}.

\subsection{Stellar properties}
\label{sec:data:properties}
Our goal was to simultaneously constrain the latitude ($\theta_f$) of the superflare on the stellar surface together with its longitude, and underlying shape. 
In order to infer $\theta_f$ from the flare light curves we required reliable information about the stars on which they occurred~(Table~\ref{tab:1}). Rotation periods ($P_{\text{rot}}$) enter both the stellar inclination calculation, and the flare modulation model itself. Spectral types are needed to determine the luminosity of the stars, and from these the flare energies ($E_f$) and flaring region sizes ($\omega$). Combining $P_{\text{rot}}$ with stellar radius ($R_*$), and projected rotational velocity $v \sin i$, we could calculate the inclination $i$ of the stellar rotation axes. Eventually, knowing $i$ allows us to break the partial degeneracy between $i$ and $\theta_f$ in the model fit.
\subsubsection{Rotation periods}
\label{sec:data:prot}
\begin{figure*}
\centerline{\includegraphics[width= 0.95\hsize]{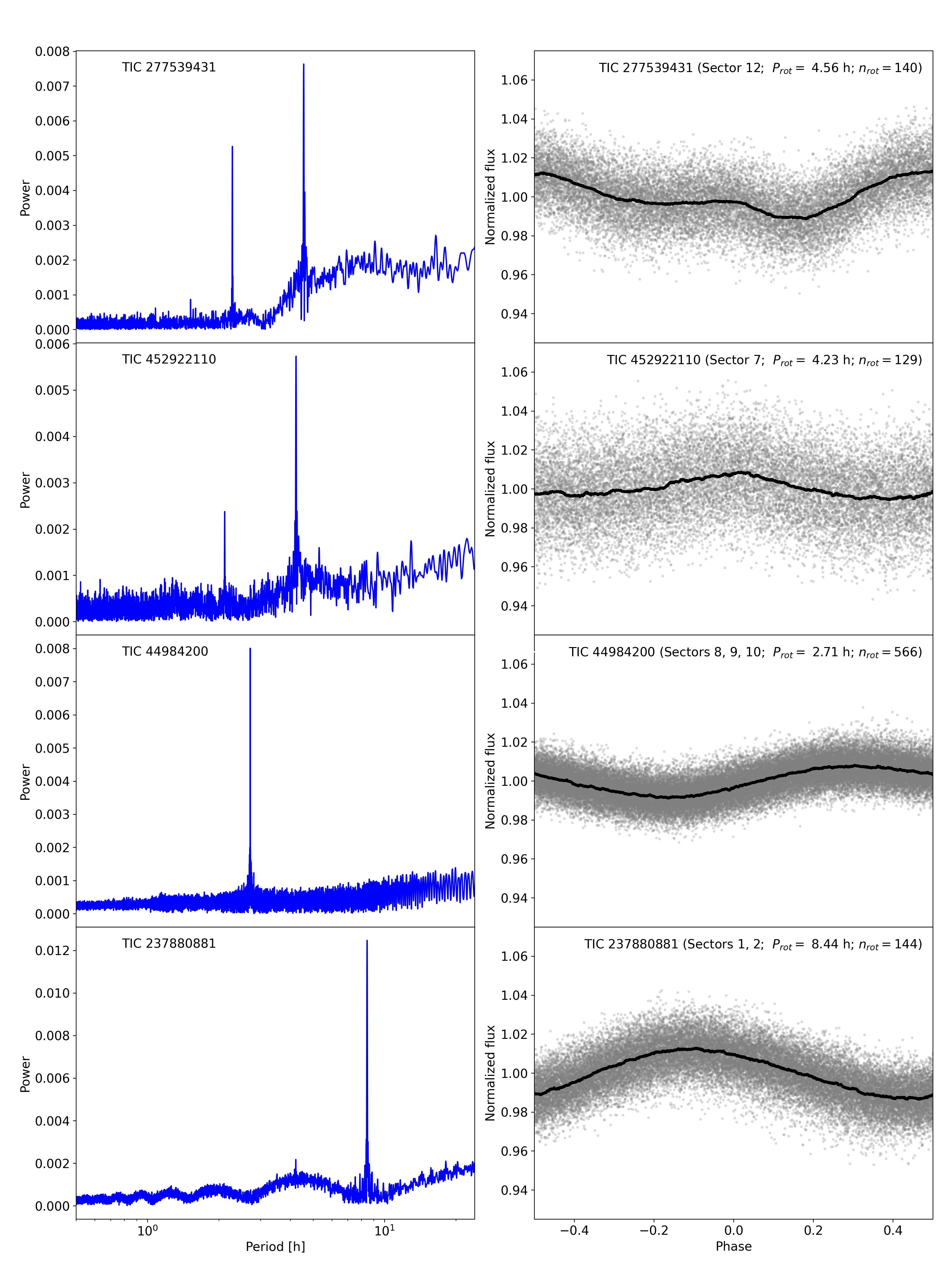}}
\caption{Left: Lomb-Scargle periodograms of the TESS light curves (\texttt{PDCSAP\_FLUX}) of the four investigated stars. Right: phase folded light curve of the same stars as on the left, folded with their rotation periods $P_{\text{rot}}$. Grey dots: Light curve flux with outliers clipped at $3\sigma$ above the noise level. Black line: Rolling median, calculated using an 800 data points window. $n_{\text{rot}}$ is the number of rotation periods covered by the data.}
\label{fig:Fig.phasefold}
\end{figure*}

We calculated the uncertainty on $P_{\text{rot}}$ by combining the results from sine fits to all available TESS light curves for each target assuming Gaussian uncertainties on the flux, and derived the posterior distributions for $P_{\text{rot}}$ using the Python Markov Chain Monte Carlo (MCMC) sampler~\texttt{emcee}~\citep{foreman_mackey2013}. The modulation was stable~(Fig.~\ref{fig:Fig.phasefold}), i.e., it did not show any sign of evolution in the phase-folded light curves of 1-3 consequent Sectors for each star. The resulting values for $P_{\text{rot}}$ were then used in the inclination calculation, and the modulation model fits.


\subsubsection{Spectral types}
We adopted spectral types from previous work~\citep{kraus2014,kirkpatrick2016,phanbao2017} that used low-resolution optical or IR spectroscopy for TIC 237, TIC 277, and TIC 449. For TIC 452, we compared a spectrum obtained from the FAST spectrogragh at the Fred Lawrence Whipple Observatory to the \citet{bochanski2007} composite spectra. For TIC 277 and TIC 449, spectral types were also consistent with the literature values when comparing the SALT spectra to those in \citet{bochanski2007}. 
We selected the nearest stars within 5 arcsec of our targets in the Two Micron All-Sky Survey~\citep[2MASS,][]{skrutskie2006} and Gaia~DR2 databases in order to obtain optical and infrared photometry and distances from parallaxes. 

\subsubsection{Stellar quiescent flux}
\label{sec:qflux}
The absolute released flare flux determines the modelled size of the flaring region in the photosphere. However, the light curves provided by TESS did not have accurately calibrated absolute fluxes. We calculated absolute quiescent fluxes in the TESS band using stellar spectrophotometry, transforming the flux in Gaia \textit{G} band to the TESS band. We first constructed representative spectra for each spectral type that covered the TESS-Gaia range ($\sim3,500-11,000\,$\AA) by combining spectroscopic templates~\citep{bochanski2007,schmidt2014a} with the IR prism spectra from the Spex Prism Library~\citep{burgasser2004, burgasser2006, burgasser2007, burgasser2008, kirkpatrick2010, burgasser2014}. We then integrated over the Gaia \textit{G} band filter response curve, and normalized the spectral energy distribution (SED) to the Gaia value. To obtain fluxes in TESS, we integrated over the response curve for TESS. The major uncertainty in the spectrophotometric fluxes was in the spectral type used to select the SED, so we assigned uncertainties to the fluxes with a difference of one spectral subclass based on typical differences in the literature, which is higher than uncertainties quoted within individual works. We then converted flux to luminosity using the Gaia distances.


\subsubsection{Stellar radii}
We calculated stellar radii based on the relationship between absolute $K_S$ magnitude and radius for M dwarfs~~\citep{mann2015}. The absolute $K_S$ magnitude was calculated from the apparent $K_S$ magnitude in 2MASS and the geometric Gaia~DR2 distance of the stars~\citep{bailerjones2018}. The uncertainties in the reported radii are based on a combination of the scatter in the magnitude-radius relationship ($\sim3$ per cent) and the uncertainties on distance and $K_S$ magnitude.

Systematic errors in the model fits may arise in the $R_*$ calculation, which would affect the derived inclination, and relative size of the flaring region. Radii that are inflated by $\sim 12-14$ per cent compared to stellar structure models have been reported for both components of the rapidly rotating late M dwarf equal mass wide binary GJ 65~\citep{kervella2016} which they attributed to youth and the suppression of convection due to strong magnetic fields. However, we determine the radii of our targets from absolute $K_S$ magnitudes, with their distance being well known from Gaia DR2 data~\citep{mann2015}. This method is self-correcting for inflation because an M dwarf that is abnormally large will also have an absolute brightness that will be larger than for a non-inflated M dwarf.

\subsubsection{Projected rotation velocity $v\sin i$}
\label{sec:vsini}
\begin{figure}
\centerline{\includegraphics[width= \hsize]{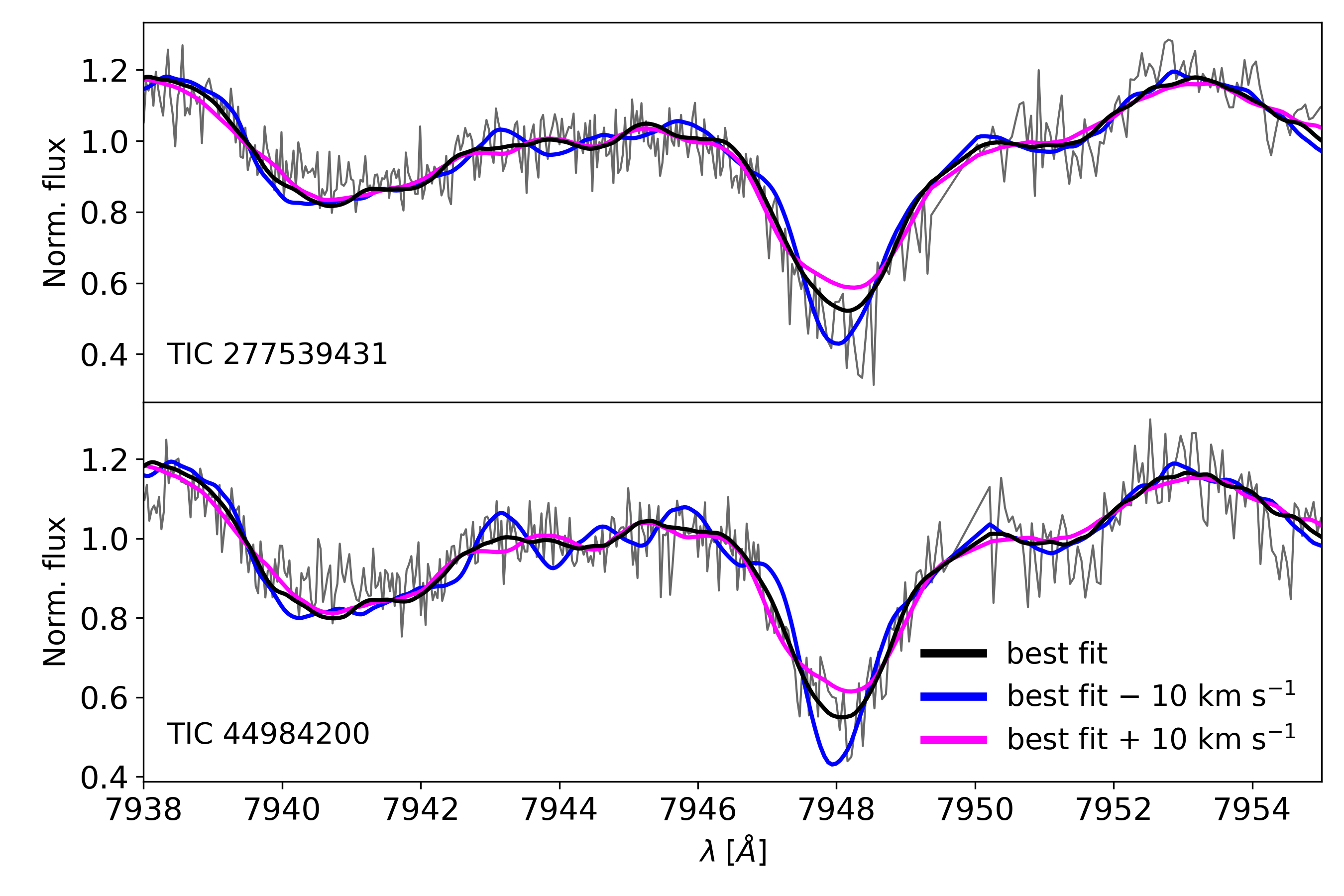}}
\caption{SALT spectra of TIC 277 (top panel) and TIC 449 (bottom panel), and rotationally broadened best-fitting template spectra. The black line is a template spectrum broadened with the best-fitting $v\sin i$ value obtained with the CCF technique. The magenta and blue lines show the same templates, but broadened with a 10 km s$^{-1}$ higher and lower $v\sin i$ than the best-fitting solution, respectively. }
\label{fig:Fig.SALT}
\end{figure}

\begin{figure}
\centerline{\includegraphics[width= \hsize]{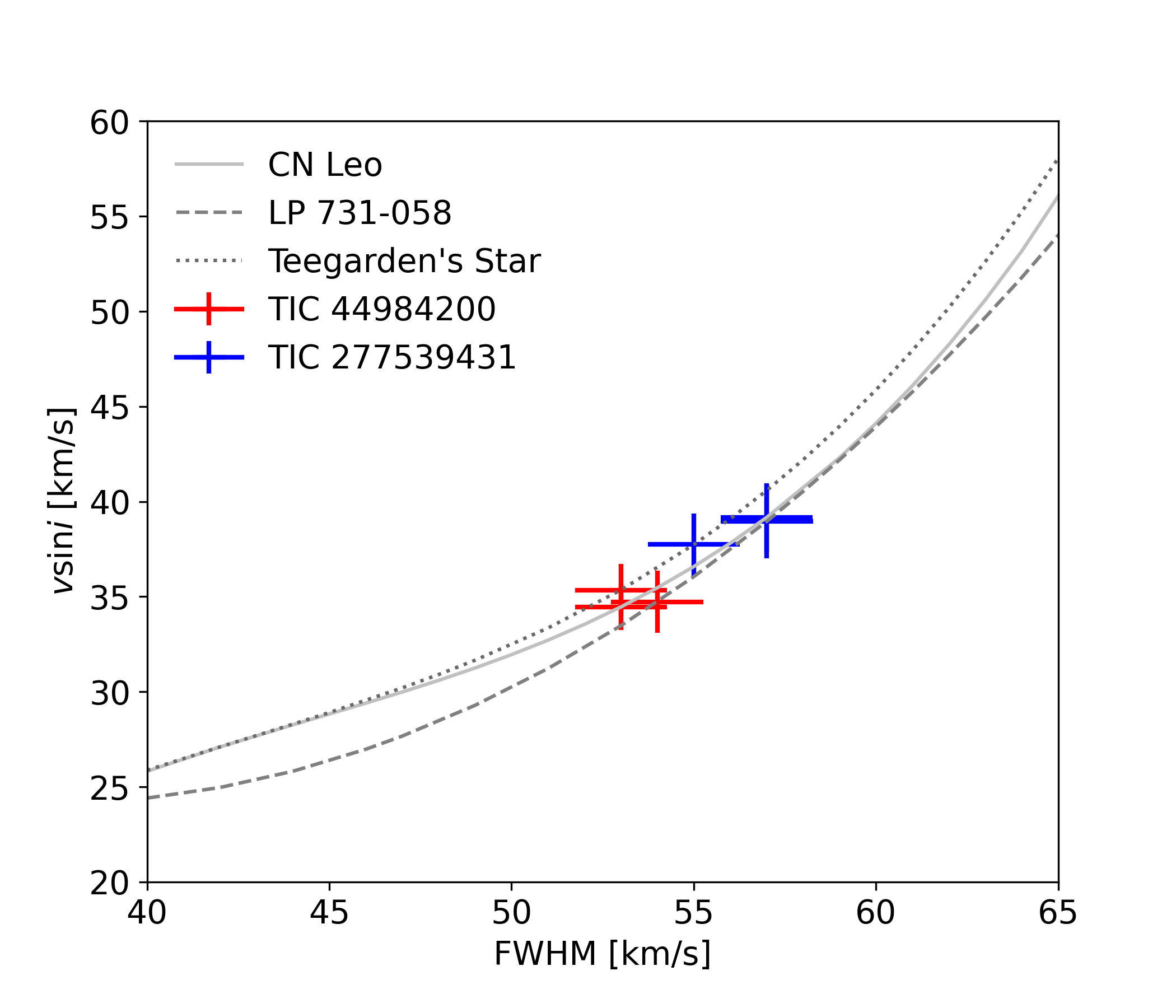}}
\caption{Projected rotational velocity $v\sin i$ derived from the SALT spectra of TIC 449 and TIC 277. Grey lines: Third order polynomial fits to the calibration functions for three template spectra. Red and blue crosses: Fitted FWHM from the cross-correlation functions of SALT spectra in the  $7,938-7,955$\AA\; region with uncertainties for TIC 449 and TIC 277. The fits indicate that the stars are rapidly rotating with $v\sin i$ of \mbox{$\sim$35 km s$^{-1}$} and \mbox{$\sim$38 km s$^{-1}$}, respectively.}
\label{fig:Fig.SM1}
\end{figure}

The inclination $i$ of the stellar rotation axis is partially degenerate with the flare latitude $\theta_f$. This ambiguity needs to be broken by an independent measurement of $i$, which we achieve by measuring the projected stellar rotation velocity $v\sin i$.

For two stars (TIC~452 and TIC~237) we used existing $v\sin i$ values~\citep{kesseli2018,kraus2014}. The other two, TIC 449 and TIC 277, did not have existing $v\sin i$ measurements in the literature. For those, we used the collected SALT spectra (see Section~\ref{sec:data:hires}) to determine the projected rotation velocity.

For the $v\sin i$ inference we used a technique that employs cross correlation functions~(CCFs) of rotationally broadened template spectra to calibrate a relation between the FWHM of the CCF and $v\sin i$~\citep{reiners2012}.
We calibrated the normalized CCFs using high-resolution optical CARMENES (Calar Alto high-Resolution search for M dwarfs with Exo-earths with Near-infrared and optical Echelle Spectrographs,~\citealt{quirrenbach2010}) spectra of CN Leo (M6), LP 731-058 (M6.5), and Teegarden's Star (M7). This was done to reduce bias in the $v\sin i$ values derived with individual templates, and to account for uncertainty in the spectral types of the SALT targets. The template stars rotate at \mbox{$v\sin i < 3$ km s$^{-1}$~~\citep{reiners2018}.}
We broadened these three template spectra using the \texttt{eniric}~~\citep{neal2019} software assuming a limb darkening coefficient $\epsilon=0.6$~\citep{claret1998} 
in the range from $1$ to $60$ km s$^{-1}$ taking $1$ km s$^{-1}$ steps. Then we measured the FWHM of the CCFs to derive a calibration function for $v\sin i$ from the broadened templates. 

In the selection of spectral regions, we started with suggestions for suitable wavelengths for low-mass stars in~\citet{reiners2018} in the $7,410-8,700\;$\AA$\;$range that overlapped with our SALT spectra. We found the region $7,938-7,955$\AA$\;$ surrounding the Rb 7948\AA$\;$ line to be best suited for the chosen technique, and the spectra collected for our target stars~(Fig.~\ref{fig:Fig.SALT}). This line is strong, shows isolated line wings, and is largely unaffected by telluric features. These properties make it sensitive to changes in $v\sin i$ in the $10-50$ km s$^{-1}$ range, while in the other inspected regions the lines were blended, or much weaker, or both.
We fit a third order polynomial to the calibration function, and used the fit to propagate the uncertainty on FWHM to the resulting $v\sin i$ values~(Fig.~\ref{fig:Fig.SM1}).
We found uncertainties on $v\sin i$ in the \mbox{$1.2-1.9$ km s$^{-1}$} range for individual pairs of templates and spectra, and no apparent trends with the spectral type of the templates, so that we treated the results as independent measurements. Averaging over the best-fit results for the pairs we arrived at uncertainties of \mbox{$0.8$ km s$^{-1}$} for TIC 449 and \mbox{$1.0$ km s$^{-1}$} for TIC 277 with Gaussian error propagation.  

\subsubsection{Stellar inclinations}
\label{sec:inclination}

\begin{figure}
\centerline{\includegraphics[width= \hsize]{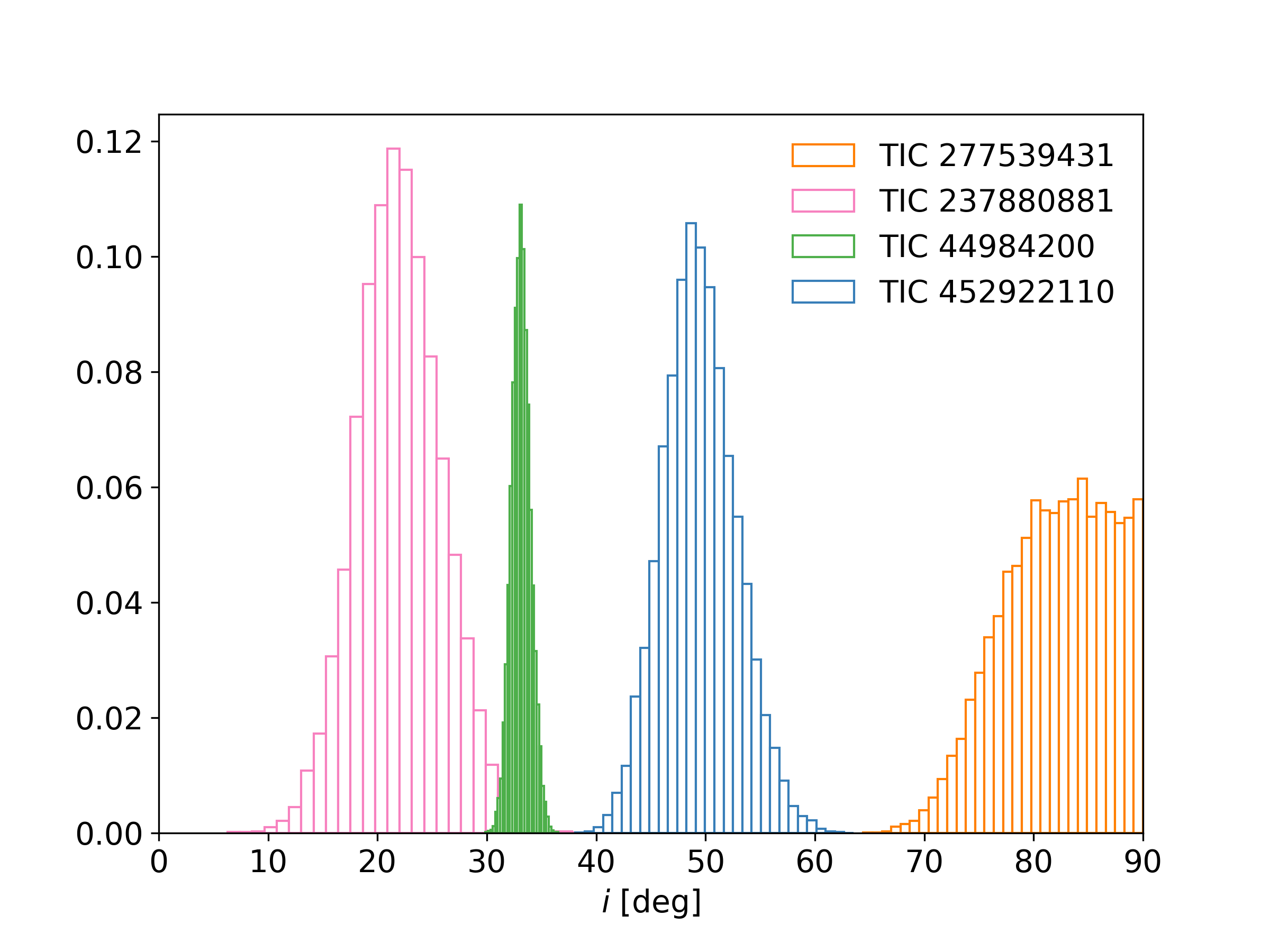}}
\caption{Posterior distributions of stellar inclination angles. Inclinations $i$ were derived from $v \sin i$, $K_s$ magnitude based stellar radii and rotation periods from TESS light curves using the method described in~\citet{masuda2020}. For all stars except for TIC 277 the uncertainties on $i$ were well approximated as Gaussian, while the latter can be approximated with a superposition of two Gaussian distributions using the convention that $i<90^\circ$. Accurate constraints on $i$ are crucial to the accuracy of the subsequent inference of the flare latitude $\theta_f$ because the morphology of a modulated flare is to some extent degenerate in $i$ and $\theta_f$.}
\label{fig:Fig.SM4}
\end{figure}


We used the stellar rotation period $P_{\text{rot}}$ and radius $R_*$ along with the $v\sin i$ to infer the stellar inclination angle $i$. To account for the statistical dependence between equatorial velocity (determined from $P_{\text{rot}}$ and $R_*$) and $v\sin i$, we used the framework of~\citet{masuda2020}. 
We used an MCMC sampler~\citep{goodman2010} to produce a posterior probability density for $\cos i$ given the measurements. We required $0<\cos i<1$, and assumed Gaussian priors on $v\sin i$, $P_{\text{rot}}$ and $R_*$. The posteriors for $i$ were non-Gaussian for the stars determined to be close to equator-on, but could be approximated by a double Gaussian fit which we subsequently used as an informative prior in the model fits~(Fig.~\ref{fig:Fig.SM4}). We give the median and $68$ per cent confidence interval as the best-fitting values and errors in Table \ref{tab:1}, using the convention that $i<90^\circ$.

\section{Methods}
\label{sec:methods}
\begin{figure*}
\centerline{\includegraphics[width=\hsize]{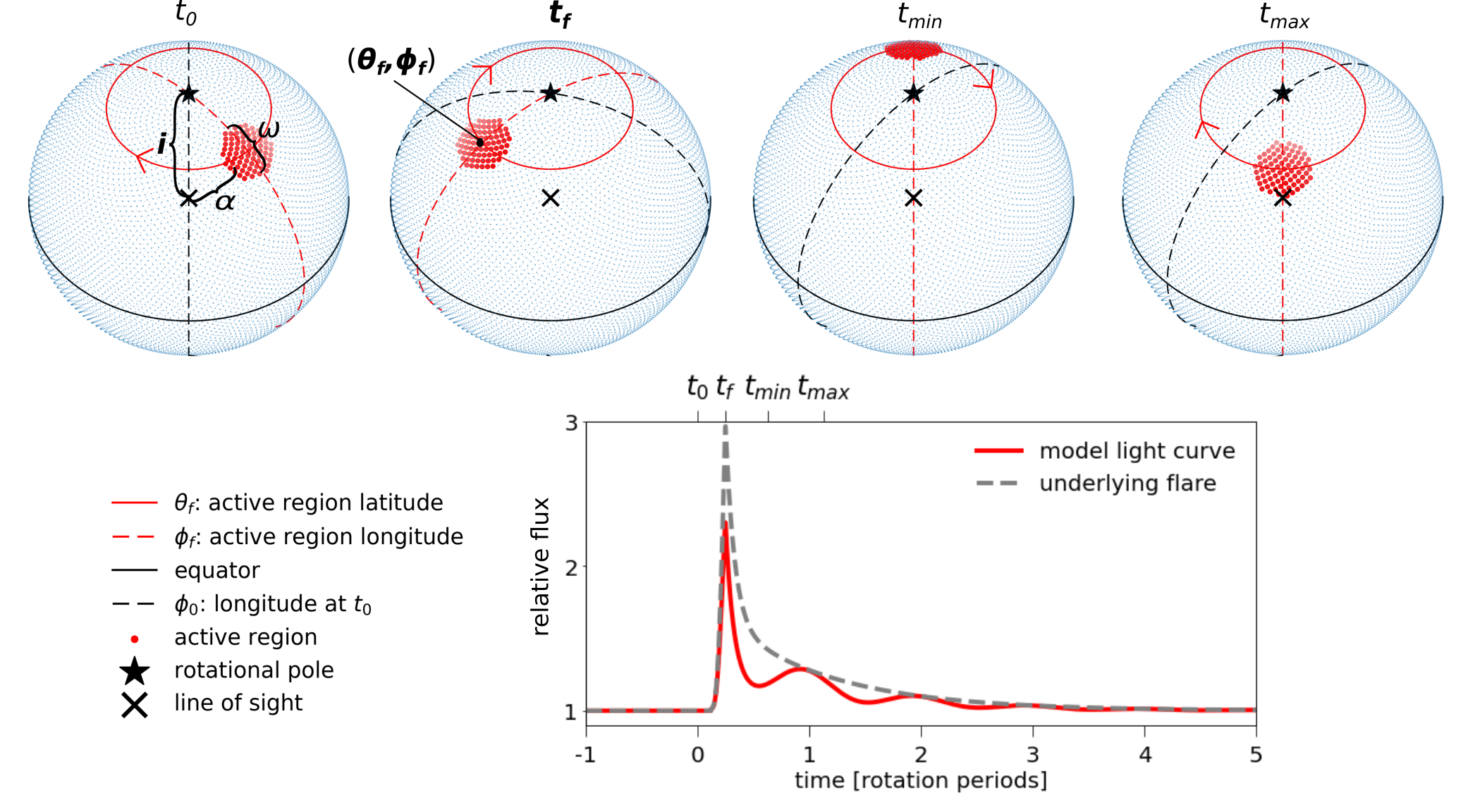}}
\caption{Flare modulation model. From left to right, the top row shows a clockwise rotating star (blue dots) with a flaring region (red dots), from the start of observation at $t_0$ to the peak flare time $t_f$, and further to the flaring region being rotated to the stellar far ($t_{min}$) and near ($t_{max}$) sides. The angular distance between the rotational pole (black star) and the intersection of the line of sight with the centre of the star (black cross) is the inclination $i$. $\alpha$ is the angular distance of the flaring region to the line of sight, and $\omega$ is the full opening angle of the circular flaring region. The depicted configuration results in the observed light curve (red line) in the bottom panel. The underlying flare model is shown as a grey dashed line in the same panel. } 
\label{fig:Fig.2}
\end{figure*}

The presence (or absence) of a periodic modulation in the flare flux observed in the optical light curves in our sample can be modeled by a bright flaring region on the stellar surface that periodically, partially or fully, rotates in and out of view (Section~\ref{sec:model}, Fig.~\ref{fig:Fig.2}). We discuss the simplifications made in the design of the model (Section~\ref{sec:methods:assumptions}), such as assuming a $10,000\,$K flare temperature, a spherical cap as the shape of the flaring region, and a universal flare time evolution parametrization. We also motivate why we neglect differential rotation in our analysis, and why we assume that multi-peaked flares are co-spatial. Our assumptions do not pose considerable limitations on our conclusions. They may, however, partly explain the cases where the model could not capture all features of the flare light curve, such as deviations from the flare template near the peak.
\subsection{The flare modulation model}
\label{sec:model}

While flaring regions on the Sun may take complicated shapes that grow, shrink, brighten and dim non-uniformly with time~~\citep{aschwanden2008}, the large flares typically observed on stars can be adequately represented in terms of the flare amplitude and the impulsive and gradual phase of the flare~\citep[e.g.,][]{guenther2020}. For our observed light curves, the integrated flux from these regions can be described as an empirical model flare~(\citealt{davenport2014},~hereafter \citetalias{davenport2014}) that erupts from a circular region and is modulated by rotation. We chose to adapt the \citetalias{davenport2014} empirical model by decoupling their $t_{1/2}$ in the first and second addend of their Eq. 4 into FWHM$_i$ and FWHM$_g$ for the impulsive and gradual flare phases, respectively. The model parametrizes the rise and the decay phases of the flare separately, as in Eqn. 1 and 4 in \citetalias{davenport2014}. In its adapted form, it reads:

\begin{eqnarray}
F_{rise} &=& 1 + 1.941\,\text{FWHM}_i - 0.175\,\text{FWHM}_i^2 \\
 &&- 2.246\,\text{FWHM}_i^3 - 1.125\,\text{FWHM}_i^4 \nonumber\\
F_{decay} &=& 0.6890\,\mathrm{e}^{-1.600\, \text{FWHM}_i}+ 0.3030\,\mathrm{e}^{-0.2783\, \text{FWHM}_g}.
\end{eqnarray}

Our (3+3+1)-parameter model was fit using an MCMC method~(Fig.~\ref{fig:Fig.1}). The first three parameters are the flare peak time $t_f$, the corresponding longitude $\phi_f$, and the latitude $\theta_f$, which define the timing of the flare and its location on the stellar surface. The other three are the relative flare amplitude $A$, FWHM$_i$, and FWHM$_g$, which determine the intrinsic shape of the light curve. The last parameter is the inclination $i$ of the stellar rotation axis, which is constrained by independent observations. The geometric modulation is determined by $i$, $\theta_f$ and $\phi_f$ that we combined to give the distance of the flaring region from the limb. The model light curve was constructed in three steps:
\begin{enumerate}
    \item Determine the size of the flaring region from the flare temperature $T_f$, and $A$, given the stellar radius $R_*$ and luminosity $L_*$.
    \item Split the flaring region into spatial elements $(\theta, \phi)$ to calculate the spatially resolved rotational modulation of the flare flux.
    \item Apply the modulation to the flare light curve and sum the spatial elements to obtain the model luminosity $L_{model}$.
\end{enumerate}

\subsubsection{Size of the flaring region}
\label{ssec:flaringregion}
We modeled the flaring region, that is, the photospheric footpoint of the flaring magnetic loop, from which the white-light emission originates, as a uniformly bright spherical cap. Its position relative to the line of sight of the observer determined the morphology of the obtained flare light curve. We described the flare emission in the optical as a \mbox{$T_f=10,000$ K} blackbody~(see Section~\ref{sec:methods:assumptions:10kk}), and determined the specific flare flux $F_{f,s}(T_f)$ of the region by integrating its spectral energy distribution within the TESS band.
The modulation is equal to the cosine of the incidence angle, known as geometrical foreshortening. The integration over the flaring region yields the maximum observable flare luminosity $L_{f,max}$, that is the flare luminosity of a flaring region centred on the line of sight~(see grey lines in Fig.~\ref{fig:Fig.1}):
\begin{eqnarray}
    L_{f,max}&=&\pi R^2 F_{f,s}(T_f) \sin^2\left(\frac{\omega}{2}\right).
    \label{eq:Lmax1}
\end{eqnarray}
$\omega$ is the full opening angle of the cap which can take values between 0 and $\pi$. In the following, all reported flare parameters (FWHM$_i$, FWHM$_g$, $A$) refer to the flare as it would appear when fixed on the line of sight. $L_{f,max}$ is also the product of $A$ with the quiescent stellar luminosity $L_{*}$ in the TESS band:
\begin{equation}
L_{f,max} = A \cdot L_{*}.
\label{eq:Lmax2}
\end{equation}
Combining Eqn.~\ref{eq:Lmax1} and~\ref{eq:Lmax2} yields the angular radius $\omega/2$ of a circular region that produces a flare with a given amplitude $A$ and temperature $T_f$ on a star with radius $R_*$ and quiescent luminosity $L_{*}$:
\begin{equation}
\omega/2 = \arcsin\left(\sqrt{\frac{A \cdot L_{*}}{\pi R_*^2 F_{f,s}(T_f)}}\right).
\end{equation}
\subsubsection{Rotational modulation and model luminosity}
The flaring region with radius $\omega/2$ centred on $(\theta_{f},\phi_{f})$ was represented by an ensemble of $N$ evenly distributed spatial elements. It was assumed to emit uniformly, so each spatial element emitted the same fraction $F_f(t)=L_f(t)/N$ of $L_f(t)$ at any given time $t$ during the flare. Before and after the flare the region emitted the average quiescent stellar flux. To model the maximum flare flux $F_f(t)$, we used the adapted~\citetalias{davenport2014} model.

For ease of calculation, we converted $t$ to phase $\hat t$ in units of radian using the rotation rate of the star $\hat t = 2\pi(t - t_{0})/P_{\text{rot}}$ where $t_{0}$ marked the start time of the light curve that we selected for the model fit, and $P_{\text{rot}}$ was the rotation period of the star.

Every spatial element emitted a modified flux $F_f(\theta,\phi,\hat t)$ of its maximum flux $F_f(\hat t)$ toward an observer at infinite distance. The modification was again given by the geometrical foreshortening with incidence angle $\alpha$:
\begin{equation}
   F_f(\theta,\phi,\hat t) = F_{f}(\hat t)\,\cos\alpha(\theta,\phi,\hat t).
 	\label{eq:lambert1}
\end{equation}
$\alpha$ was the distance from the intersection point $O$ of the line of sight with the centre of the star to every spatial element ($\theta$, $\phi$)
\begin{equation}
    \alpha = \arccos\left(\sin\theta \cos i + \cos\theta \sin i \cos(\phi - \phi_0 - \hat t)\right),
    \label{eq:alpha}
\end{equation}
where $\phi_0$ was the longitude that faced the observer at $t_{0}$. 

To determine when a spatial element was behind the limb we calculated the visible rotation fraction $D$, i.e., the fraction of each stellar rotation during which the spatial element is visible to a distant observer. Applying the spherical law of cosines to the triangle between $O$, a point on the limb at a given latitude, and the rotational pole of the star we obtained
\begin{equation}
    D = \frac{1}{\pi}\arccos\left(-\tan\theta \tan\left(\frac{\pi}{2}-i\right)\right).
    \label{eq:lambert2}
\end{equation}
Using $D$ and $\phi_0$ we defined a step function $d(\hat t)$ that was 1 when the spatial element was visible, and 0 when it was hidden.
Combining $d$ with $F_f(\theta,\phi,\hat t)$ we obtained the modulated flux of a spatial element in the flaring region
\begin{equation}
    F_{model}(\theta,\phi,\hat t) = F_f(\theta,\phi,\hat t)\,d(\theta,\phi,\hat t).
    \label{eq:lambert3}
\end{equation}
Finally, we calculated the model luminosity $L_{model}$ as a function of $\hat t$ by summing up all spatial elements. For $N\to\infty$, $L_{model}$ becomes the surface integral over the flaring region.
In our fits, we used \mbox{$N=100$}, motivated by the time resolution of TESS light curves in cases where the modeled flare is modulated with a high gradient, that is, when the impulsive flare peak is moving in or out of view at the limb. In all other cases, $N \approx 10$, or even $N=1$ (for flares located close to the pole at low $i$) can be sufficient.
\subsubsection{Fitting the model to the observed light curve}
First, we masked the flare and fit the dominating starspot modulation and long-term trends in the light curve using Gaussian Process (GP,~\citealt{rasmussen2006gaussian}) regression with \texttt{celerite2}~\citep{celerite1,celerite2}. \texttt{celerite2} is a software that provides a library of covariance functions, or kernels, to use for GP regression on one-dimensional data sets like time series, and more specifically, light curves. Following~\citet{celerite1}, we chose a mixture of two kernels, called SHOTerm, which are derived starting from the dynamics of stochastically driven damped simple harmonic oscillators (SHOs) as a general model for stellar variability. The first is a RotationTerm, which is itself a mixture of two SHOTerm kernels with periods $P_{\text{rot}}$ and $P_{\text{rot}}/2$, and used to model the rotational variability. The second is a single SHOTerm that covers all other variability that may be present in the light curve. We obtained posterior distributions on the parameters using the same wide Gaussian priors ($\sigma=2$) for all fits except for the prior on the rotation period in the RotationTerm kernel. There, we used the period $P_{\text{rot}}$ obtained from the Lomb-Scargle periodogram as a prior (see Sections~\ref{sec:data:prot} and~\ref{sec:data:flares}). We then subtracted the prediction for the light curve in the time interval of the masked flare to obtain the residual flare light curve to which we subsequently fit our model.

In a second step, we constructed a likelihood function assuming that the uncertainty in flux (quadratic addition of uncertainty from GP fit and \texttt{PDCSAP\_FLUX\_ERR}, termed "original uncertainty" from here on) was Gaussian. Our priors were uniform over the physically permissible ranges ($>0$ for the flare parameters, $[0^\circ,360^\circ]$ for the longitude of the flare, and $[-90^\circ,90^\circ]$ for $\theta_f$) for all parameters but $i$, for which we adopted empirical priors~(see Section~\ref{sec:inclination}). Using the MCMC method~\citep[\texttt{emcee},][]{foreman_mackey2013} we determined the posterior distributions of the flare peak time $t_f$; the intrinsic flare properties FWHM$_i$, FWHM$_g$,  and $A$; the location of the flaring region, $\theta_f$ and $\phi_f$; and an updated distribution of the inclination $i$.


\subsection{Model assumptions and their motivation}
\label{sec:methods:assumptions}
\subsubsection{Flare emission in the optical is a 10 000 K blackbody}
\label{sec:methods:assumptions:10kk}
We assumed that the observed red-optical emission of our flares stems from a blackbody component at effective temperature \mbox{$T_f\approx 10,000$\,K,} as is common in both the Sun and other low-mass stars~\citep{hawley1992, kretzschmar2011, kowalski2013,kowalski2018}. Hotter temperatures have occasionally been reported for M dwarf superflares in the past~\citep{robinson2005,froning2019}. Typically, the short, impulsive phase can be hotter than $\sim 10,000$\,K, but cools rapidly to $\lessapprox 10,000$\,K where it emits during the longer, gradual phase~\citep{fuhrmeister2008, howard2020}. In our model, $T_f$ primarily affects the size of the flaring region. A large region will be more smoothly modulated than a small bright kernel. If an initially hotter temperature with subsequent rapid cooling was present, our model would overestimate the size of the flaring region in the first few minutes of the flare decay, but the protracted gradual phase that characterizes long-duration superflares would be largely unaffected.
\subsubsection{The flaring region is a spherical cap}
We also assumed that the flaring region was a simple spherical cap on the stellar surface. White-light flares on the Sun take more complex shapes in general, such as pairs of non-circular footpoints, or crescent-shaped ribbons~\citep{hudson2006}. The morphologies of active regions that are associated with the strongest flares are also the most complex~\citep{sammis2000}. In particular, non-circular shapes may cause departures from our model when the region moves across the stellar limb, but less so when the region is fully in view. Moreover, effects like saturation of flare loops might violate the assumption that the flaring region is flat~\citep{heinzel2018, jejcic2018}. However, unlike in solar observations, we can neglect limb darkening effects~(see Appendix~\ref{appendix:limb}).

\subsubsection{Flares follow the same time evolution parametrization}
\label{sec:methods:timeevo}
Flares often display a complex time evolution that does not follow the~\citetalias{davenport2014} empirical template. The majority of the largest flares are complex. Based on the $\sim 6,800$ flaring events that \citet{guenther2020} detected in TESS M dwarf light curves, out of $10$ per cent of the flares with the highest amplitudes ($A>0.24$) more than $52$ per cent were best fit with superpostions of two or more \citetalias{davenport2014} flare templates, while in the other $90$ per cent this fraction was only $\sim 20$ per cent.

Some flares cannot be resolved into superpositions of multiple classical flares (``unusual'' flares). While we expect about one out of four flares to be complex~\citep{hawley2014,davenport2014,guenther2020}, unusual flares are rare with about $1$ per cent incidence rate~\citep{davenport2014}. 

The \citetalias{davenport2014} model was derived from the flares found in Kepler light curves of GJ 1243, a frequently flaring dM4e star. We based our flare model on the empirical parametrization derived for this star. However, if the time evolution of the flare and the geometry of the flaring region deviate from our assumptions, we expect this to mostly affect the impulsive phase of the flare. To avoid overfitting the gradual decay phase of the flare which reveals more about the rotational modulation than the impulsive phase due to its relatively long duration, we made two adjustments. First, we chose to decouple the time scales for the impulsive and gradual phases. This is consistent with models that attribute different physical processes to the impulsive and gradual phases of flares~\citep{benz2010}. Second, we increased the uncertainties on the flux above $A/2$ in each of the flares to $20$ per cent of the flare-only flux, as shown in the bottom panels in Fig.~\ref{fig:Fig.1}. The number was an ad hoc choice aiming to permit relatively strong variability in the amplitude while preserving the overall shape of the flare.

\subsubsection{Differential rotation effects are negligible}
\label{sec:methods:assumptions:diffrot}
Differential rotation may in principle cause systematic errors in the inclination estimates. Assuming the predicted differential rotation $(\Omega_{equator}-\Omega_{pole})/\Omega_{bulk}\sim 0.02$~\citep{yadav2015} for fully convective stars, derived inclinations would increase by about $0.5^\circ-5^\circ$. Assuming complete degeneracy between $i$ and $\theta_f$, the flare latitudes would \textit{move poleward} accordingly. The poleward shift would be at the high end of the estimate for flare latitutes that were initially close to the pole, and at the low end for flares near the equator. However, fully convective stars do not appear to reach such strong differential rotation, instead observations indicate typical values about one order of magnitude lower~\citep{barnes2017}. Moreover, the degeneracy is only partial, because one can, for instance, distinguish a pole-on from an equator-on flare for a relatively broad range of inclinations, but less so for smaller differences in latitude (compare, for instance, the left- and rightmost columns in Fig.~\ref{fig:Fig.SM2}). Overall, as long as differential rotation is Sun-like, i.e. rotation is slower at the pole than at the equator, the real flare latitudes will be a bit closer to the poles than what we derive in this work.

Another effect of differential rotation would cause the rotation period derived from the light curve to differ from the rotation period at the latitude of the flare. The input period would be either too long or too short, resulting in a poor fit if an otherwise strong rotational modulation of the flare light curve is present. 
However, we observed no lag between between the minima of modulation and the spot induced variability. 
\subsubsection{Flare footpoints are co-spatial with the loci of magnetic field emergence}
One caveat concerns the interpretation of the flare latitude as the locus of magnetic field emergence (see Section~\ref{sec:intro}, and references therein): We cannot exclude that the region of white-light emission footpoints could have wandered across the stellar surface in the time between emergence and flare eruption. The distance between where the magnetic field concentration is at the time of flux emergence and at the time of white-light emission will depend on the duration of the build-up phase of the flare, and the conditions of the sub-surface plasma. Our stars show stable starspot configurations over timescales of multiple weeks to months which suggests that large scale spot evolution is very slow, but we cannot exclude more dynamic conditions on smaller scales. 
\section{Results}
\label{sec:results}
\begin{figure*}
\centerline{\includegraphics[width=\hsize]{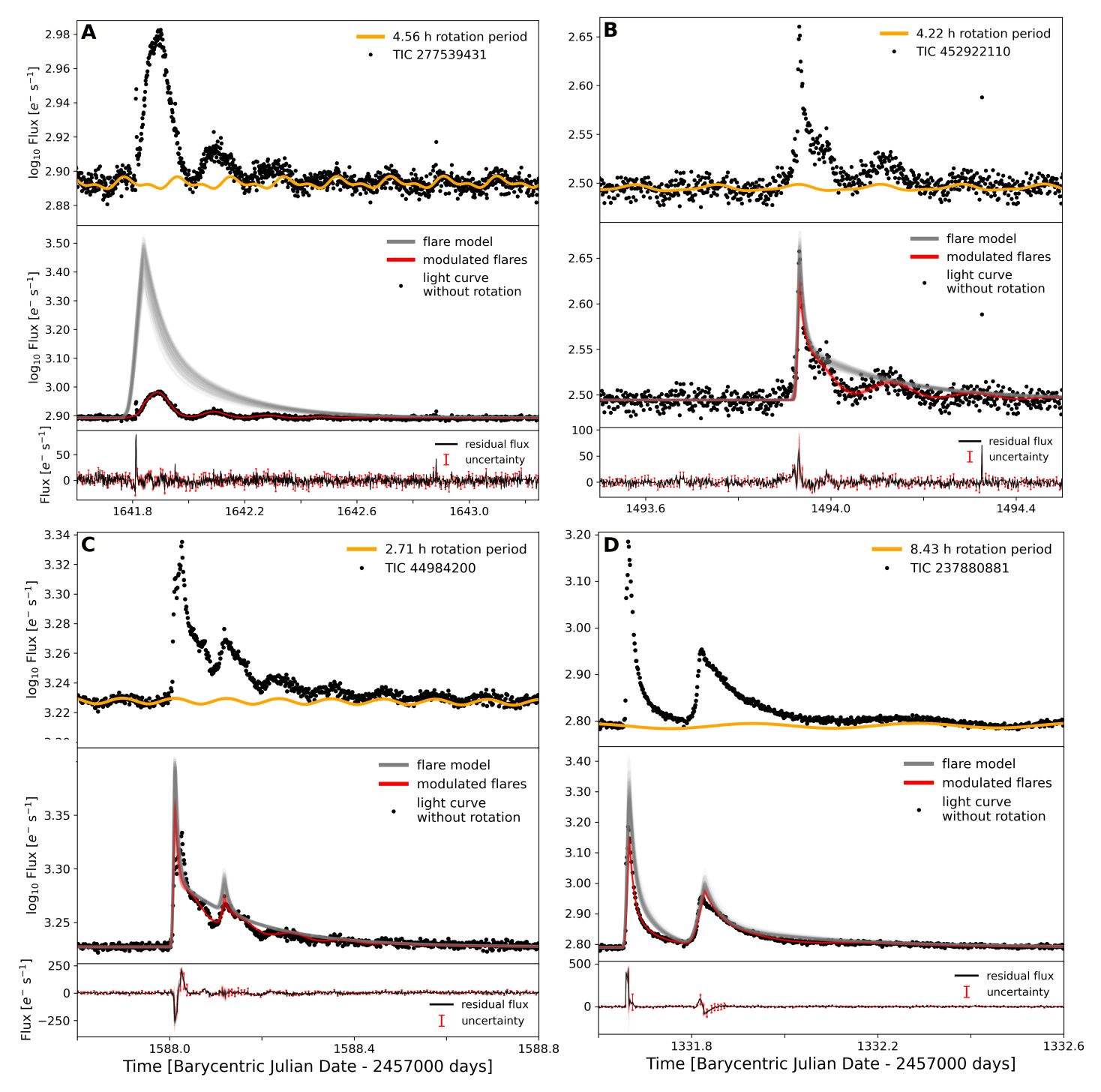}}
\caption{Giant flares on rapidly rotating stars. Each of the subfigures A-D shows one superflare on a fully-convective fast-rotating star that was visible for more than a full stellar rotation period. \textit{Top panels.} Observed flare in the TESS light curve (\texttt{PDCSAP\_FLUX}, black dots). Quiescent light curve fit incorporating long term trends and rotational variability from starspots (orange line). \textit{Mid panels.} TESS light curve with rotational variability of the non-flaring photosphere removed (black dots). Samples from the posterior distribution of the fit to the modulated flare curve (red lines). Samples from the distribution of underlying flares without rotational modulation (grey lines). \textit{Bottom panels.} Residual flux after subtracting rotational modulation and the model fit (black line). Flux uncertainty (red error bars). The flares in the subfigures A-C show clear modulation in phase with stellar rotation, while the flares in subfigure D remain nearly unaffected. We also constrain the longitude at which the flare occurs within a few percent uncertainty, e.g., in subfigure A, the longitude of the flare peak is about $230^\circ$, implying that it must have occurred behind the limb given the high inclination of the rotation axis.}
\label{fig:Fig.1}
\end{figure*}

\begin{table*}
\caption{Properties, observations, and best-fitting results for stars with superflares. Stars with two flares are indexed 1 and 2 for the first and second flare in chronological order. Values in parentheses are the differences to the $84$th (upper) and $16$th (lower) percentiles, or else the 1-$\sigma$ uncertainty. SpT: spectral type. $K_S$: 2MASS magnitude. $d$: Gaia distance. $P_{\text{rot}}$: rotation period of the star. $v \sin i$: projected rotation velocity of the star. $R_*/R_\odot$: stellar radius measured in solar radii. $i$: updated posterior of the stellar inclination on the informative prior derived from $v\sin i$, $P_{\text{rot}}$ and $R_*$. $\log_{10} E_f$: flare energy released in the TESS band. $A$: amplitude of the underlying flare as a fraction of quiescent flux. FWHM$_i$ and FWHM$_g$: impulsive and gradual full-width-at-half-maximum flux of the respective flare phases. $\theta_f$: latitude of flaring region on the stellar surface.}
\begin{tabular}{l|cccccc}
\hline\hline
{} &                                            TIC 277 &                                            TIC 452 &                        TIC 449 (2-flare) &                        TIC 237 (2-flare) \\
\hline
SpT                       &                                                 M7 &                                                 M7 &                                       M6 &                                       M5 \\
$K_S$ (mag)               &                                  $9.666 \pm 0.024$ &                                 $10.865 \pm 0.021$ &                        $9.268 \pm 0.021$ &                       $10.636 \pm 0.023$ \\
$d$ (pc)                  &                                   $13.70 \pm 0.11$ &                                   $22.03 \pm 0.06$ &                         $11.10 \pm 0.03$ &                         $46.01 \pm 0.14$ \\
$P$ (min)                 &                                $273.618 \pm 0.007$ &                                $254.094 \pm 0.054$ &                      $162.694 \pm 0.007$ &                      $506.145 \pm 0.020$ \\
$v \sin i$ (km s$^{-1}$)  &                                     $38.6 \pm 1.0$ &                                     $29.7 \pm 1.5$ &                           $34.8 \pm 0.8$ &                           $14.4 \pm 2.6$ \\
$R_*/R_\odot$             &                                  $0.145 \pm 0.004$ &                                  $0.137 \pm 0.004$ &                        $0.142 \pm 0.004$ &                        $0.275 \pm 0.008$ \\
$i$ (deg)                 &                  $87.0\left(^{+2.0}_{-2.4}\right)$ &                  $49.2\left(^{+3.3}_{-3.3}\right)$ &        $33.1\left(^{+0.9}_{-0.9}\right)$ &        $21.3\left(^{+3.7}_{-3.9}\right)$ \\
$\log_{10} E_{f,1}$ (erg) &            $34.473\left(^{+0.067}_{-0.084}\right)$ &            $33.531\left(^{+0.024}_{-0.021}\right)$ &  $33.355\left(^{+0.006}_{-0.006}\right)$ &  $34.599\left(^{+0.034}_{-0.027}\right)$ \\
$A_1$                     &                $2.61\left(^{+0.45}_{-0.47}\right)$ &                $0.47\left(^{+0.04}_{-0.04}\right)$ &      $0.50\left(^{+0.01}_{-0.01}\right)$ &      $2.33\left(^{+0.24}_{-0.19}\right)$ \\
FWHM$_{i,1}$ (min)        &                 $103.5\left(^{+5.7}_{-5.4}\right)$ &                  $21.0\left(^{+5.1}_{-3.4}\right)$ &        $14.5\left(^{+1.4}_{-2.3}\right)$ &        $18.2\left(^{+0.9}_{-0.8}\right)$ \\
FWHM$_{g,1}$ (min)        &                  $85.1\left(^{+2.4}_{-2.4}\right)$ &                  $74.8\left(^{+6.3}_{-5.3}\right)$ &        $64.3\left(^{+1.1}_{-1.0}\right)$ &        $16.9\left(^{+0.6}_{-0.5}\right)$ \\
$\log_{10} E_{f,2}$ (erg) &                                                ... &                                                ... &  $31.994\left(^{+0.066}_{-0.075}\right)$ &  $34.670\left(^{+0.026}_{-0.013}\right)$ \\
$A_2$                     &                                                ... &                                                ... &      $0.08\left(^{+0.01}_{-0.01}\right)$ &      $0.60\left(^{+0.04}_{-0.03}\right)$ \\
FWHM$_{i,2}$ (min)        &                                                ... &                                                ... &        $25.5\left(^{+4.0}_{-4.0}\right)$ &        $63.0\left(^{+2.5}_{-2.7}\right)$ \\
FWHM$_{g,2}$ (min)        &                                                ... &                                                ... &         $3.9\left(^{+1.8}_{-2.3}\right)$ &        $93.6\left(^{+2.0}_{-2.0}\right)$ \\
$\theta_f$ (deg)          &                  $80.9\left(^{+0.5}_{-0.6}\right)$ &                  $63.1\left(^{+3.5}_{-3.7}\right)$ &        $71.9\left(^{+1.1}_{-1.1}\right)$ &        $55.2\left(^{+5.1}_{-6.1}\right)$ \\
\hline

\end{tabular}

\label{tab:1}
\end{table*}

\begin{figure}
\centerline{\includegraphics[width=\hsize]{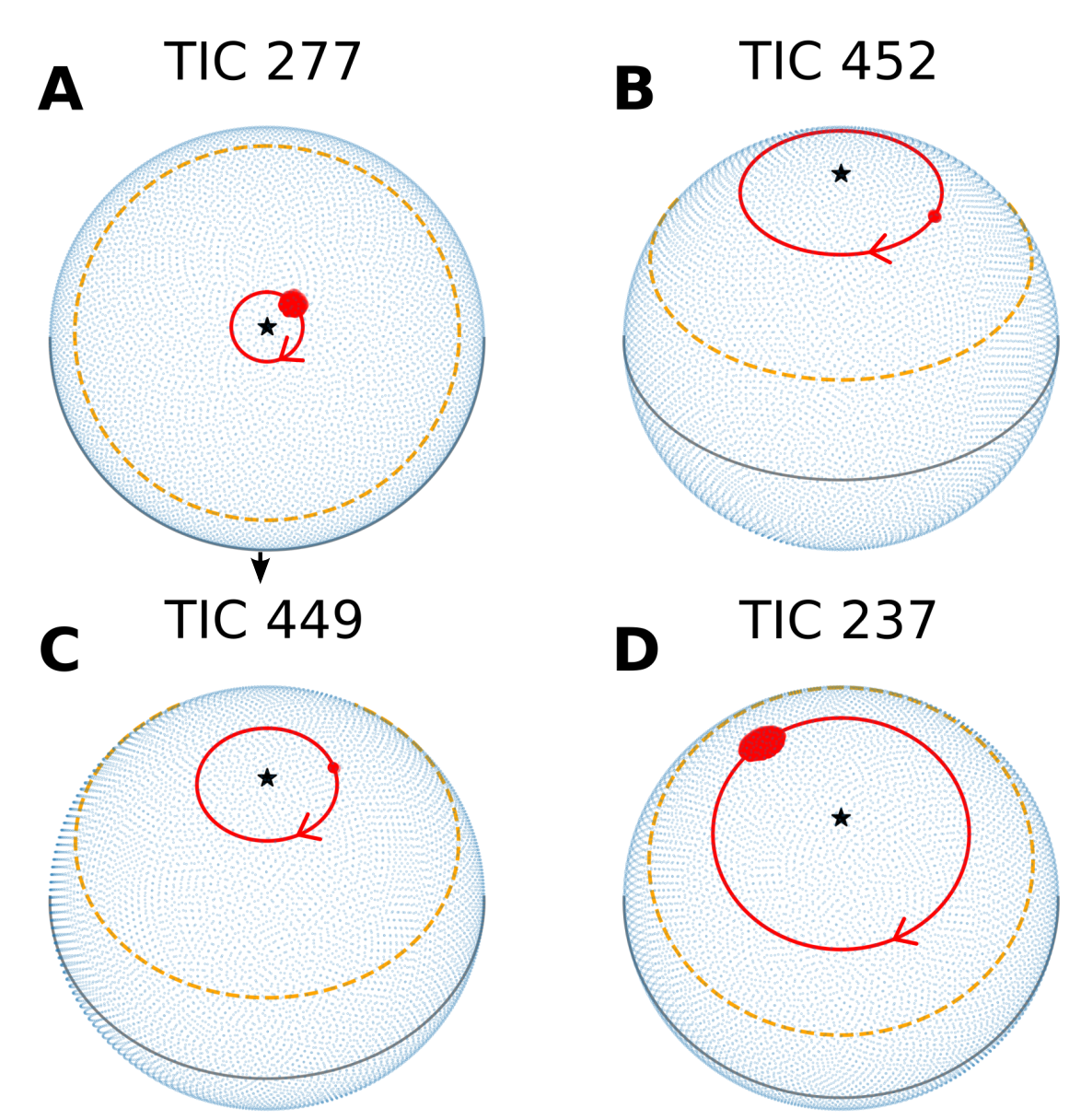}}
\caption{Loci of giant flares at flare peak time on the stellar surface, the subpanels correspond to those in Fig.~\ref{fig:Fig.1}. Red line: flare latitude. Red spot: active region, shown to scale. Orange dashed line: maximum typical solar flare latitude ($30^\circ$). Black star: rotational pole of the star. Grey line: stellar equator. In sub-panel A, the flaring region is behind the limb. In sub-panels C and D the longitudes are those of the first flare on each star. TIC 277 (sub-panel A) is observed close to equator-on. For better visibility, we tilted the rotation axis by $90^\circ$. The observer's line of sight is indicated as black arrow.}
\label{fig:Fig.1b}
\end{figure}
Polar spots have been observed in fast-rotating young Sun-like stars~\citep{isik2018}. However, on the present-day Sun, sunspots, and the flares they are associated with, occur in a belt around the equator, and almost never at latitudes above $30^\circ$. We found all our selected flares at significantly higher latitudes around and above $55^\circ$.

We found four candidates with long-duration superflares, and shorter measurable rotation period $P_{\text{rot}}$: TIC 277~(Sector 12), TIC 449~(Sector 10), TIC 452 (Sector 7), and TIC 237 (Sector 1). They rotate so quickly ($P_{\text{rot}}=2.7-8.4$ h) that the observed white-light superflares last longer than a full rotation period on each star. Three of the four stars display strong flux modulation during the superflare decay \textit{in phase with the stellar rotation}. This can most straightforwardly be explained by rotational modulation, with the flaring region rotating in and out of view on the stellar surface.

In Fig.~\ref{fig:Fig.1} we show all investigated superflare light curves. The shape of the modulation contains information on the latitude of the flaring region: the combination of flare latitude and the inclination angle of the stellar rotation axis determine how long a flaring region is visible as the star rotates. Similarly, unless the star is viewed strictly pole-on, the absence of flare modulation also contains latitude information, indicating that the flare occurs close enough to the rotational pole for the modulation to be indistinguishable from the noise, and that the pole is visible all the time; this is exploited for TIC~237 which displays much more subtle but still measurable rotational flare modulation.

We found that all four stars displayed their giant flares at high latitudes $\geq 55^\circ$. In Table \ref{tab:1}, we give the results for the marginalized posterior distributions of the model parameters, and in Fig.~\ref{fig:Fig.1b}, we illustrate the inferred active regions loci at flare peak time on the stellar surface. The full posterior distributions can be viewed in detail in Appendix~\ref{appendix:mcmc}.

\subsection{Model setups: single and double-flare, original and increased uncertainty around the flare peaks}
\label{sec:results:setups}
The stars TIC~277 and TIC~452 both showed single modulated flares. The flares we detected on TIC 449 and TIC 237 were complex because they were better fit with a superposition of two flares than with a single flare~(see Section~\ref{sec:methods:timeevo}).

TIC~237 displayed two flares in close sequence, both without apparent rotational modulation. We fitted this light curve under the assumption that both flares occurred at the same location, which is common on the Sun~\citep{uchida1968, wheatland2006, torok2011} and proposed for other stars as well~\citep{vida2016, gizis2017b}. TIC~449 displayed excess flux indicating a secondary flare during the second rotation of the main event that was, however, less clear than in the case of TIC 237. A model with a single underlying flare for TIC 449 is unable to reproduce the shape of the light curve through rotational modulation, in particular around the secondary peak. A model with two co-spatial flare events reproduces the light curve better.

Since flares often display complex evolution in the impulsive phase and the flare peak, we accomodated possible deviations from the flare template by increasing the uncertainty to $20$ per cent of the excess flux around the peak (above $A/2$, see Section~\ref{sec:methods:timeevo}) for all flares except for TIC~277 where the flare peak most likely occurred behind the limb, and is therefore not observed (bottom panels in Fig.~\ref{fig:Fig.1}). This is motivated by the fact that our model is most sensitive to the (absence of) rotational modulation in the gradual decay phase. Deviations in the short impulsive phase, which our model cannot capture, however, can cause over-fitting of the template because they are more likely intrinsic to the flare than to effects of geometrical foreshortening.

We could improve or achieve consistent modelling of the modulated decay phase of the flare in all cases when we increased the uncertainties. In Appendix \ref{appendix:altfits}, we show the alternative fits with original uncertainties from the TESS light curves~(Fig. \ref{fig:Fig.SM9}, \ref{fig:Fig.SM6} A and B, \ref{fig:Fig.SM5}). While we preferred the fits shown in Fig.~\ref{fig:Fig.1}, the inferred latitudes were overall high even in the solutions with higher amplitudes and stronger systematic deviations in residual flux ($>48^\circ$,~see Table~\ref{tab:2}).
\subsection{Star by star: model fits, alternative and disfavoured setups}
\label{sec:results:alternatives}
\subsubsection{TIC 277}
TIC 277 (WISEA J105515.71-735611.3) is an M7 dwarf 13.7 pc away that rotates with a 4.56 hour period and is seen nearly equator-on.
The superflare occurred at $\theta_f=\left(80.9^{+0.5}_{-0.6}\right)^\circ$ so that the flaring region was only briefly hidden from view during each rotation. 

The flare peak occurred while the region was facing away from the observer (Fig.~\ref{fig:Fig.1} A) at phase $\phi_f=0.64\pm 0.01$, and was a clear single flare event, so that we did not fit the model with alternative setups. Note that the center of the flaring region becomes visible around $\phi\sim 0.70$, while other parts of it stay behind the limb longer (at lower latitudes) or shorter (at higher latitudes).

The uncertainty on $\theta_f$ is small despite the relatively high uncertainty on inclination~(Fig.~\ref{fig:Fig.SM4}). This is due to the sensitivity of the modulation model to flares that occur simultaneously close to the observed limb and to the rotational pole~(see posterior probability distributions in Fig.~\ref{fig:Fig.277mcmc}). For the flare in TIC 277, the flare light curve morphology alone constrains the latitude to very high values even without knowing the inclination. 

One possible caveat is the flare temperature. If the flare is visible throughout almost the entire rotation, which is the case for TIC 277, $i$ can be considered a lower limit on $\theta_f$, unless the flaring region is very large. The flaring region size depends on the flare temperature, which we here assumed to be $10,000$ K. If $T_f$ were only $6,500$ K, the observed flare could be placed at about $70^\circ$ latitude. However, this temperature would be more typical for less energetic solar flares~\citep[see Appendix in][and references therein]{castellanos2020} than for stellar superflares, which are usually hotter~(see Section~\ref{sec:methods:assumptions:10kk}). Similarly, a higher temperature could place the flaring region even closer to the poles.
\subsubsection{TIC 452}
TIC 452 (2MASS J08055713+0417035) is an M7 dwarf that has literature \mbox{$v\sin i=(29.7\pm 1.5)$~km~s$^{-1}$}~\citep{kesseli2018}, and a Gaia DR2 distance of $22$ pc.
A full rotation period of this star is 4.22 h, and the rotation axis is viewed at an inclination angle of about 49$^\circ$.
Its chromospheric activity places TIC 452 among active fast-rotating low-mass stars with saturated H$\alpha$ luminosity~\citep{newton2017}.
Both the single flare fit with original~(Fig. \ref{fig:Fig.SM9}) and the preferred fit with increased uncertainties above $A/2$~(Fig.~\ref{fig:Fig.1} B) consistently implied a flare latitude of $\approx 63^\circ$.

In the remaining two stars, the simple model had to be adjusted to a more complex flare morphology.
\subsubsection{TIC 449}
TIC 449 (SCR J0838-5855) is an M6 dwarf 11.1 pc away rotating at $P_{\text{rot}}=2.71$ h. 
The flare on TIC 449 was visible for almost five rotation periods. While the flux appeared modulated, the flaring region never fully rotated out of view.
The light curve strongly suggests the presence of a secondary flare. However, the burst in the second maximum could also stem from other complex flaring behaviour in the wake of the large event. We therefore fitted both a single- and a co-spatial two-flare solution to the light curve with increased and original uncertainties above $A/2$~(see Section~\ref{sec:results:setups}). The first and second events are indexed accordingly in Tables~\ref{tab:1}~and~\ref{tab:2}.

The fit with the least systematic residual flux in the important gradual decay phase was achieved in the two-flare fit with increased peak uncertainties~(Fig.~\ref{fig:Fig.1} C) which indicated that the flares occurred at high latitudes at $(72\pm 1)^\circ$. Because of the contamination by complex behaviour in the secondary maximum the results of the fits were inconsistent with each other, but yielded high latitudes regardless: $(86\pm 1)^\circ$ for the two-flare fit with original uncertainties~(Fig. \ref{fig:Fig.SM6} A), $(83\pm 1)^\circ$ for the single flare fit with original uncertainties~(Fig. \ref{fig:Fig.SM6} B), and $(70\pm 1)^\circ$ for the single flare with increased uncertainties~(Fig. \ref{fig:Fig.SM6} C).
While a secondary flare is a likely scenario, other conceivable explanations include a flare region shape that deviates substantially from our circular simplification, or a cluster of $>2$ flares. 
\subsubsection{TIC 237}
\begin{figure}
\centerline{\includegraphics[width=\hsize]{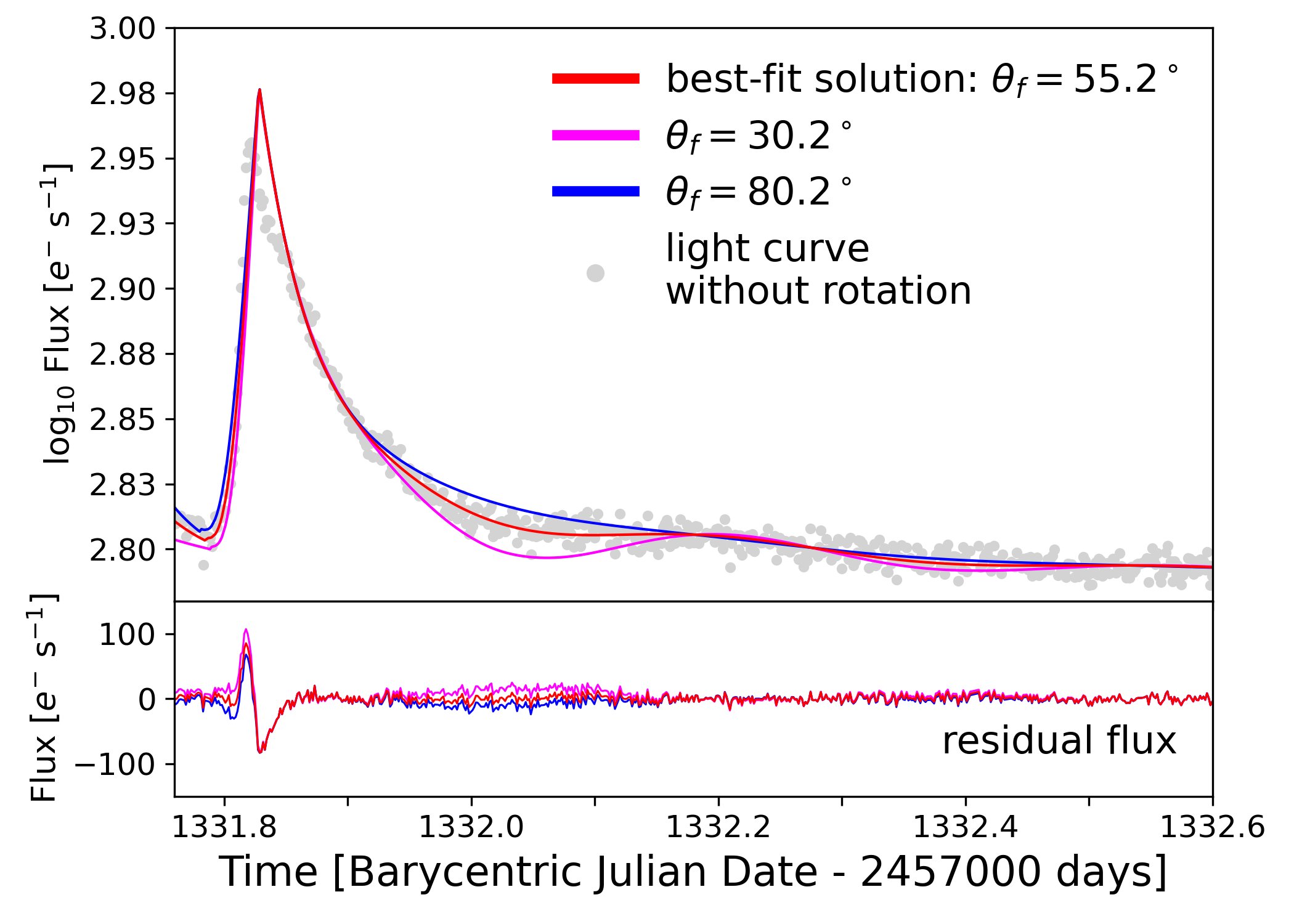}}
\caption{Close-up on the second event in the light curve of TIC 237. Upper panel. Grey dots: flare light curve with rotational variation removed. Red line: best-fit solution (see Table~\ref{tab:1}). Magenta and blue lines: best-fit solution with FWHM$_i$ and $A$ adjusted to match the peak and impulsive decay phase of the flare while decreasing and increasing the flare latitude $\theta_f$ by $25^\circ$, respectively. Lower panel: residual flux for the three different models, color-coded as in the upper panel.}
\label{fig:Fig.237latitudes}
\end{figure}

Finally, TIC 237 (2MASS J01180670-6258591) was our earliest type star, an M5 dwarf at a distance of 46 pc with literature $v\sin i = (14.4 \pm 2.6)$~km~s$^{-1}$~\citep{kraus2014}.
TIC 237 is a probable member of the Tucana-Horologium association~\citep{ujjwal2020}, a young nearby moving group with age estimates spanning $4-45$ Myr~\citep{ujjwal2020, bell2015, kraus2014}.

The systematic deviation of the fit with original uncertainties, which implied $\theta_f\sim 48^\circ$, was stronger in the decay phase~(Fig.~\ref{fig:Fig.SM5}) than that with increased uncertainties on both peaks~(Fig.~\ref{fig:Fig.1} D), which suggests a flaring region latitude of $\left(55^{+5}_{-6}\right)^\circ$.

In TIC 237, the flare modulation was very subtle yet measurable, in particular, in the decay phase of the second event. We show a close-up in Fig.~\ref{fig:Fig.237latitudes}, in which we placed the flaring region at $25^\circ$ higher and lower latitutes relative to the best-fit solution, respectively. The observed modulation appears as an intermediate between stonger modulation at lower latitudes, and vanishing modulation closer to the pole. We cannot rule out that at least a fraction of the observed modulation is intrinsic to the flare. If all variation was intrinsic, the flare would reside on the pole. In a less extreme case, if there was a mixture of intrinsic variation and rotational modulation that gives the appearance of more rotational modulation than is actually present, the derived latitude would still be biased to lower values. The opposite case, where rotational modulation is masked by intrinsic variation, is less likely to occur: While variation that mimicks rotational modulation can manifest as modulation with many different flare latitudes and longitudes, a variation that masks rotational modulation has much fewer degrees of freedom. We therefore suggest to give more credence to the lower error quoted for TIC 237 than the upper error. The same reasoning applies, albeit to a lesser degree, to the other three flares.

\section{Discussion}
\label{sec:discussion}

The consistently high latitudes of the superflares studied in this work are significant, even though the sample size is small. If superflares were uniformly distributed across the stellar surface, the probability of finding all four superflares above $55^\circ$ would be about $0.1$ per cent, i.e. very unlikely, provided that our detection method is not biased towards finding preferentially high latitude flares (Section~\ref{sec:discussion:detection}). If flares occurred at all latitudes $\theta_f$ equally, and all inclinations $i$ were equally likely to be observed, purely geometric considerations imply that as many flares would be observed in the equatorial strip $[-26.4^\circ, 26.4^\circ ]$ as in the complementing northern and southern polar caps taken together.  

Superflares that last longer than a full rotation period of the star on fully convective M dwarfs are rare, with about one event per star per 120 years of continuous observation based on our sample, and distinct from the quasiperiodic pulsations phenomenon~(Section~\ref{sec:discussion:pulsate}). Yet each event is an opportunity to probe the magnetic nature~(Section~\ref{sec:discussion:magnetic}) and the space weather conditions (Section~\ref{sec:discussion:space}) on these cool, ubiquitous stars. 
\subsection{Observing polar vs. equatorial rotationally modulated superflares}
\label{sec:discussion:detection}

\begin{figure*}
\centerline{\includegraphics[width=12cm]{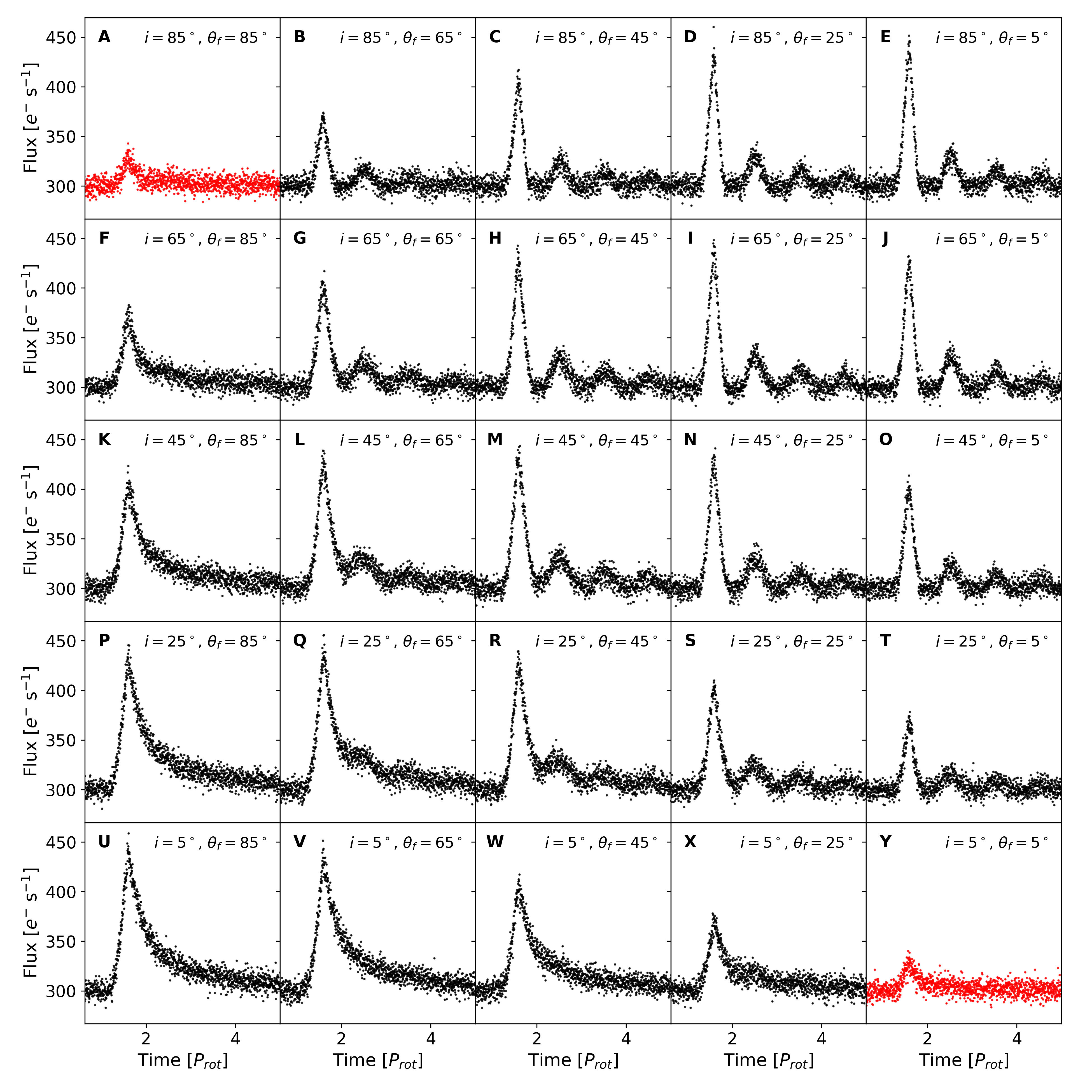}}

\caption{Light curves of a subset of 25 simulated multiperiod flares. Setup: Relative amplitude $A=0.5$, FWHM$_i=$FWHM$_g=4 P_{\text{rot}}$, and $2$ per cent Gaussian noise on a fictitious star with varying inclination $i$ and flare latitude $\theta_f$. Black scatter: recovered events (sub-panels B--X). Red scatter: missed events (sub-panels A and Y).}

\label{fig:Fig.SM2}
\end{figure*}

To test if our selection mechanism disfavours equatorial flares, we analysed the recovery efficiency of our pipeline on a grid of simulated flares that we injected into light curves with typical noise properties. To produce the synthetic light curves, we used our model to produce flares on a uniform grid of latitudes $\theta_f$ and inclinations $i$. We fixed the stellar properties and flare FWHM values (FWHM$_i \equiv$  FWHM$_g$), and varied the underlying flare amplitude $A$ and peak time $t_f$ to capture the effects of photometric noise levels and the visibility of the flare peak on the detection probability and recovered properties. We show an example of 25 flares from this grid of model flares that occur at different combinations of $i$ and $\theta_f$ in Fig.~\ref{fig:Fig.SM2}. The majority of flares shown in this example would have been recovered, except for events that remain attenuated for the entire duration of the event because they occurred close the limb (red scatter). These can be both polar flares seen at high inclinations (Fig.~\ref{fig:Fig.SM2}~A), and equatorial flares seen at very low inclinations (Fig.~\ref{fig:Fig.SM2}~Y). Overall, our automated flare detection method (Section~\ref{sec:data:flares}) did not show any bias towards polar over equatorial flare locations.

\subsection{Quasiperiodic pulsations are ruled out}
\label{sec:discussion:pulsate}
Rotational modulation is the preferred scenario for our observations because the period of the modulation matches the rotation period of the star. A different type of flare modulation, magnetohydrodynamic oscillations of plasma following strong flares, has been seen on the Sun~\citep{nakarianov1999,aschwanden1999} and some other stars~\citep{lopezsantiago2016,pillitteri2014}. Such quasiperiodic pulsations have been mainly observed in coronal X-ray emission or in chromospherically active individual spectral lines~\citep{srivastava2008, nakariakov2001, mitrakraev2005} with periods ranging from a few and up to about 20 minutes~\citep{inglis2012}.
White-light flare oscillations have been measured in a number of Sun-like and low-mass stars, but the detected periods were predominantly between 5 and 45 min, and none exceeded 100 min~\citep{pugh2016}, remaining shorter than the shortest modulation period we found.
In contrast, the flare modulation we report here is of the order of several hours, and is observed in white-light. 

\subsection{Stellar dynamos, magnetic field emergence, and polar spots}
\label{sec:discussion:magnetic}
Large polar spots have long been suspected to exist on fully convective fast rotators, based on indirect methods such as (Zeeman) Doppler Imaging (DI or ZDI), which map spectral modulations to stellar surfaces~\citep{morin2008,morin2010,barnes2017,kuzmychov2017}. Moreover, a number of simulations predict that strong magnetic fields emerge near the poles in fully convective fast rotators~\citep{gastine2013,yadav2015,weber2016}. 

The body of knowledge on magnetic field configurations of rapidly rotating, fully convective stars is still small, not least because the targets are faint and atomic lines that are typically employed for these methods in earlier type stars fade out and the spectrum grows dominated by molecular signatures. However, magnetic fields of up to 7 kG have been reported from fully convective fast rotators~\citep{shulyak2017}, far above the equipartition values suggested by solar-type dynamo models~\citep{pevtsov2003}. The average value is about 5 kG as derived from molecular spectropolarimetry in a sample of nine M1 to M7 dwarfs~\citep{afram2019}. Among the later M5-M7 dwarfs, cool spots appeared to occupy the majority of the stellar atmosphere, and at a greater depth than in the earlier types.

In ZDI maps of fully convective dwarfs with spectral types \mbox{$\geq$ M4}, a wide range of magnetic filling factors was observed~\citep{see2019}. This can be interpreted as manifestation of a bimodal dynamo with either strong bipolar or weak multipolar fields~\citep{morin2010}. 
The strong bipolar branch is associated with fast rotation, a stable spot configuration, and little differential rotation~\citep{gastine2013}, suggesting that the stars in our sample belong on this branch, and rotate as near-solid bodies.

Studies on individual stellar systems indicate the presence of both branches. V374 Peg (M4V, $P_{\text{rot}}=0.446$ d~\citep{newton2016}), showed no dominating polar spot, and weak contrast small spots on its surface at most~\citep{morin2008peg}. GJ 65 A and B, two rapidly rotating fully convective dwarfs with similar properties, revealed very different spot configurations~\citep{barnes2017}.
While GJ 65 A appeared to have spots in a band around $30-40^\circ$, but also larger polar spots at $50-85^\circ$, observations of GJ 65 B are more consistent with extended spots at intermediate latitudes ($50-56^\circ$) and no spots above $70^\circ$. GJ 791.2A (M4.5V, $P_\mathrm{rot} = 0.257$ d, \citealt{newton2016}) showed spots at all latitudes, but circumpolar spots were larger than those closer to the equator. The spot structure of LP 944-20 (M9V) could be reconstructured with a dominating large spot at $79^\circ$, and marginal detections of smaller spots at lower latitudes~\citep{barnes2015}.

Finally, \citet{berdyugina2017} reconstructed a magnetic field loop from giant flare-like bursts of the fast-rotating~(2.84 h,~\citealt{hallinan2008}) M8.5 dwarf~\citep{deshpande2012} LSR J1835+3259, one of which appeared to be modulated by rotation. Lacking a precise stellar rotation axis inclination, they could not derive the latitude of the loop's footpoints. However, they recovered the geometry of the loops, and found that if the bursting region occupied $\sim 3$ per cent of the stellar surface ~\citep{hallinan2015} -- comparable to the sizes of flaring regions in our sample -- its magnetic field strength could reach 10 kG. 

In this context, our results are further evidence that in fully convective, rapidly rotating dwarfs, strong magnetic field concentrations may indeed emerge at latitudes that are closer to the rotational pole than to the equator, where they create active regions that can erupt in giant flares. 

\subsection{Implications for stellar space weather}
\label{sec:discussion:space}
Our findings are relevant to exoplanets; several thousand of such planets outside of the Solar System have been discovered up to now~\citep{perryman2018}. Planets residing in the habitable zones around their host stars, that is,  orbital distances that allow for liquid water on the planetary surface, are of prime interest. The habitable zone is located close to the host star in the case of fully convective M dwarfs, enabling efficient detection of temperate planets, in comparison to more massive host stars.
In order to characterize habitability of temperate planetary surfaces, the longevity and composition of exoplanetary atmospheres needs to be taken into account. 

Numerical simulations have shown that flares and coronal mass ejections, which are directly associated with flares in the solar case~\citep{webb2012}, can erode or alter exoplanetary atmospheres, and cause loss of exoplanetary oceans~\citep{shields2016,tilley2019}.
Our results suggest that the largest flares on such fully convective stars preferably occur at high latitudes. This decreases the flare energy that is received at the planetary habitable zone orbit, but will less affect X-ray and (E)UV emission~(Section \ref{sec:discussion:space:attenuate}). 
Coronal mass ejections are thought to accompany stellar superflares~\citep{drake2013,odert2020}. However, in contrast to flares, they are only indirectly and rarely observed in stars to date~\citep{moschou2019}. If they can escape the strong magnetic field of the star~\citep{alvaradogomez2018} and propagate at large angles to the planetary orbital plane they would pose a lesser risk to their planets' environments than predicted under the assumption of a Sun-like low latitude distribution~(\citealt{tilley2019}, Section 2.4., and our Section~\ref{sec:discussion:space:cme}). These considerations assume that the planetary orbital plane and the stellar equator are roughly aligned, which is not well understood for late-type systems~(Section \ref{sec:discussion:space:alignment}). 

\subsubsection{The impact of flares decreases with increasing flaring region latitude}
\label{sec:discussion:space:attenuate}
\begin{figure}
\includegraphics[width= \hsize]{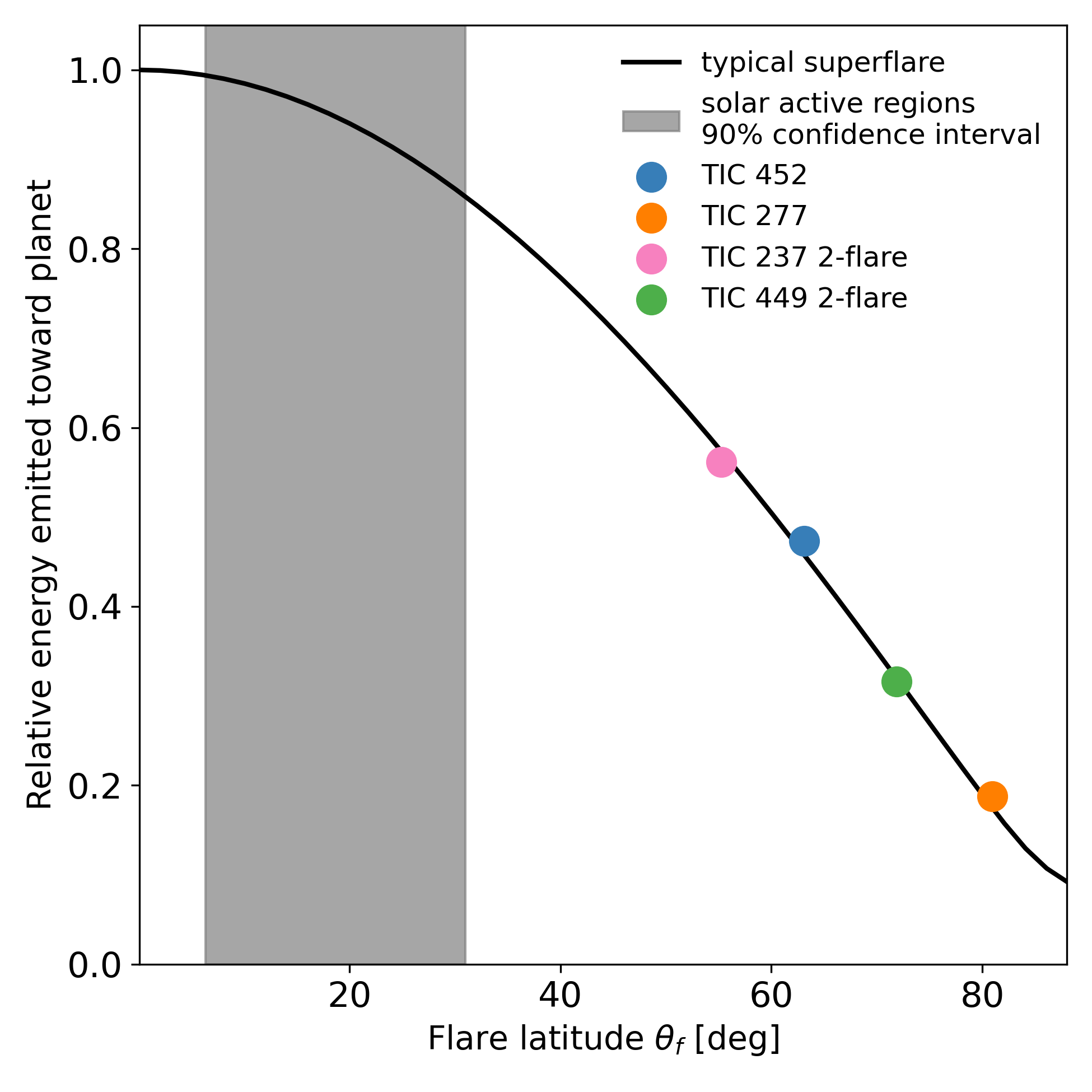}
\caption{Attenuation of optical continuum emission from a flat flaring region as a function of flaring region latitude. Relative optical flare energy emitted in the ecliptic plane of a spin-orbit aligned star-planet system as a function of flare latitudes $\theta_f$ for a typical superflare (black line). Colored dots indicate the relative energies at the $\theta_f$ derived for the stars in this study. The thermal emission of large flares at high latitudes is significantly attenuated. The grey shaded region indicates the latitudes of 45 solar active regions that comprised less than $0.5$ per cent of all active regions but produced about half of all X-class flares in solar activity cycles 21-23~\citep{chen2011}.} 
\label{fig:Fig.SM3}
\end{figure}

With the present results, we have the opportunity to estimate the relative flux received by exoplanets, that are potentially hosted by these or similar stars, as a function of flare latitude. All else equal and assuming spin-orbit alignment, placing the flares in our sample at $45^\circ$ or $80^\circ$ latitude, the optical flux emitted in the direction of a distant observer in the orbital plane decreases by $\sim 30$ and $\sim 85$ per cent compared to the same event located near the equator. The exception is the first flare in the light curve of TIC 237 that, when placed at  $80^\circ$, is visible for a larger fraction of the stellar rotation period due to the relatively large size of the active region. We show the typical relative attenuation of optical flux as a function of flare latitude in Fig.~\ref{fig:Fig.SM3}.

This comparison assumes that compact bright kernels in the stellar photosphere contribute the majority of the blackbody radiation to the flare events. This is true for most solar flares, but it is possible for superflares with particularly large flare region sizes ($\gtrapprox 2.5$ per cent of the total stellar surface) to saturate the density in the flare loop arcades in the gradual decay phase such that they become optically thick and contribute significantly to, and may even become the dominant source of, white-light emission~\citep{heinzel2018, jejcic2018}. Flare emission from a volume with considerable height, as it is proposed by~\citet{heinzel2018}, instead of a flat slab on the stellar surface may explain the residual flux in our model fits in the impulsive phase because geometrical foreshortening cannot be applied to an extended 3D loop. Yet the good overall agreement between our model and data in the gradual phase suggests that the flat circular kernel is a satisfactory approximation for the purpose of accurate latitude inference.

However, the more energetic (E)UV and X-ray emission is produced at higher altitudes on the Sun. Spectropolarimetric observations of a large flare on a late M dwarf suggest a similar stratification for the stellar case~\citep{berdyugina2017}. Furthermore, stellar coronae are optically thin, meaning that the X-ray photons originating from them are not subject to the self-shadowing effects that optical emission experiences near the stellar limb. The X-ray flux received by the planets may therefore be much less attenuated than the optical continuum appears in Fig.~\ref{fig:Fig.SM3}. Ultimately, an accurate assessment of flare radiation impact on habitable zone exoplanets will have to include the combined effects of flaring region location, strong magnetic fields, high density, and large pressure gradients in the compact atmospheres of late type stars and brown dwarfs on the 3D emission distribution of superflares.
\subsubsection{Coronal mass ejections and energetic particle events}
\label{sec:discussion:space:cme}
The majority of X-class flares on the Sun are associated with coronal mass ejections (CMEs,~\citealt{compagnino2017}). A similar correlation is expected for stellar superflares~\citep{odert2020}, but direct CME observations are missing to date~\citep{crosley2018}, perhaps due to differences in the spectral signature between solar and stellar CMEs~\citep{alvaradogomez2019,alvaradogomez2020}. Solar energetic particle (EP) events typically carry a few percent of the energy of the CME they are associated with~\citep{aschwanden2017}. Extrapolating the power law relation between X-ray flux and solar EP fluence to stellar superflares is poorly constrained so that stellar events may contain one to five orders of magnitude more energy than the largest EP events on the Sun~\citep{herbst2019}. If stellar EP events associated with the most powerful flares are ejected from the star at high latitudes, as suggested by our observed flares, they may typically not collide with an exoplanet's atmosphere, rendering their effect on habitability smaller than for equatorial flares. However, such particles coming from CMEs and EPs might also be deflected towards the equatorial plane by a strong stellar dipole field~\citep{kay2019}. The impact of particle events also hinges upon their ability to leave the stellar magnetosphere. In M dwarfs, there is evidence that CMEs and EPs may be suppressed by the large overarching dipole field that these stars often possess~\citep{alvaradogomez2018,fraschetti2019}.
\subsubsection{Key uncertainty: Spin-orbit alignment}
\label{sec:discussion:space:alignment}
All of the above assumes that exoplanetary orbital and their host stars' rotation axes are aligned.  
While there is statistical evidence for spin-orbit alignment in Kepler exoplanet hosts~\citep{mazeh2015}, cases of misaligned systems suggest that alignment is less common than one would expect from protoplanetary disk formation~\citep{winn2015, bourrier2018, louden2020, hjorth2021}. For small planets around M dwarfs, however, only few measurements exist, all of which indicate spin-orbit alignment~\citep{hirano2020b,addison2020,palle2020,stefansson2020}. Spin-orbit alignment in low-mass star systems with Earth-like planets is therefore not yet well understood, which ultimately limits our ability to assess the influence of flare emission direction on exoplanetary space weather. 

\section{Summary and conclusions}
\label{sec:summary}
In a systematic analysis of fully convective stars observed with TESS, we detected four stars that displayed giant flares which were modulated in brightness by the stars' rapid rotation. The exceptional morphology of the modulation allowed us to directly localize these flares between $55^\circ$ and $81^\circ$ latitude on the stellar surface.
Our findings are evidence that strong magnetic fields tend to emerge close to the rotational poles of fast-rotating fully convective stars, and suggest a reduced impact of these flares on exoplanet habitability.

Our search encompassed the first two years of TESS observations, during which one hemisphere per year was observed. By summer 2021, TESS will have completed its second full-sky scan that will give us the opportunity to put the conclusions of this work to test with more detections of long duration superflares.

The strength of our method lies in its complementarity to spectroscopy and spectropolarimetry not only with respect to the type of scientific information used to infer the stars' magnetic properties, but also regarding the time scales probed. Flares trigger sudden and drastic changes in the local magnetic field configuration. Space-based time series photometry can capture its effects with a few minutes or less resolution. In contrast, ground based (Z)DI reconstructs larger scale changes on time scales of hours and longer. 

This work is a step towards understanding spatially constrained flare evolution on stars other than the Sun. Together with multiwavelength flare studies~\citep{kowalski2013,guarcello2019,namekata2020,maehara2020} that provide insight into the propagation of flare events through the stellar atmosphere we may soon see the first fully empirical spatio-temporal flare reconstructions on low mass stars. Such models would help us understand the so far elusive relation between stellar flares, coronal mass ejections, and energetic particle events, and better assess the impact of stellar magnetic activity on the space weather environment of exoplanet systems.

\section*{Acknowledgements}
The authors would like to thank Uwe Wolter for his thorough reading and constructive criticism that markedly improved the quality of this work. EI acknowledges support from the German National Scholarship Foundation. KP acknowledges support from the German \textit{Leibniz Community} under grant P67/2018. JRAD acknowledges support from the DIRAC Institute in the Department of Astronomy at the University of Washington. The DIRAC Institute is supported through generous gifts from the Charles and Lisa Simonyi Fund for Arts and Sciences, and the Washington Research Foundation. This paper includes data collected with the TESS mission, obtained from the MAST data archive at the Space Telescope Science Institute (STScI). Funding for the TESS mission is provided by the NASA Explorer Program. STScI is operated by the Association of Universities for Research in Astronomy, Inc., under NASA contract NAS 5-26555. This study was enabled by the TESS guest observer program G022164 led by JSP, ERN, JRAD, SJS, and Adam Burgasser. This research has benefitted from the SpeX Prism Library, maintained by Adam Burgasser. We are thankful to Andrew Vanderburg for providing the code used to calculate the inclination probability distributions.
\\
Some of the observations reported in this paper were obtained with the Southern African Large Telescope (SALT). This publication makes use of data products from the Two Micron All Sky Survey, which is a joint project of the University of Massachusetts and the Infrared Processing and Analysis Center/California Institute of Technology, funded by the National Aeronautics and Space Administration and the National Science Foundation. This work has made use of data from the European Space Agency (ESA) mission {\it Gaia} (\url{https://www.cosmos.esa.int/gaia}), processed by the {\it Gaia} Data Processing and Analysis Consortium (DPAC, \url{https://www.cosmos.esa.int/web/gaia/dpac/consortium}). Funding for the DPAC has been provided by national institutions, in particular the institutions participating in the {\it Gaia} Multilateral Agreement. Based on data from the CARMENES data archive at CAB (INTA-CSIC). This research also made use of the Python packages numpy, pandas astropy, and the cross-match service provided by CDS, Strasbourg.

\section*{Data availability}

TESS light curves are publicly available through the Mikulski Archive for Space Telescopes (\url{https://mast.stsci.edu/portal/Mashup/Clients/Mast/Portal.html}). The Gaia data used in this work are publicly available through the Gaia archive (\url{https://gea.esac.esa.int/archive/}). Data from the 2MASS Point Source Catalogs is accessible through the Centre de Donnees astronomiques de Strasbourg (CDS) via the VizieR Service (\url{http://vizier.u-strasbg.fr/cgi-bin/VizieR?-source=II/246]}). CARMENES spectra are publicly available through the CARMENES data archive (\url{http://carmenes.cab.inta-csic.es/gto/jsp/reinersetal2018.jsp}). SpeX Prism Library is available through \url{http://www.browndwarfs.org/spexprism}. Reduced SALT spectra ara available via Zenodo, under DOI:\url{10.5281/zenodo.4332142}. Low-resolution spectra obtained from the FAST spectrogragh at the Fred Lawrence Whipple Observatory (used for spectral typing of TIC 452) will be made available online through the Dataverse (preparation of data release in progress, to be completed prior to acceptance).  Python code including the flare modulation model, model fitting scripts, Gaussian Process regression scripts, scripts that were used to produce the figures and tables in this manuscript is publicly available through GitHub (\url{https://github.com/ekaterinailin/MalachiteMountains}). The code used to calculate the inclination probability distributions was provided by A. Vanderburg for use in this work and is available by request to the authors and with permission from A. Vanderburg.



\bibliographystyle{mnras}
\bibliography{mnrasbib} 




\appendix

\section{Alternative and disfavoured fits to the light curves}
\label{appendix:altfits}
\begin{figure}
\centerline{\includegraphics[width=\hsize]{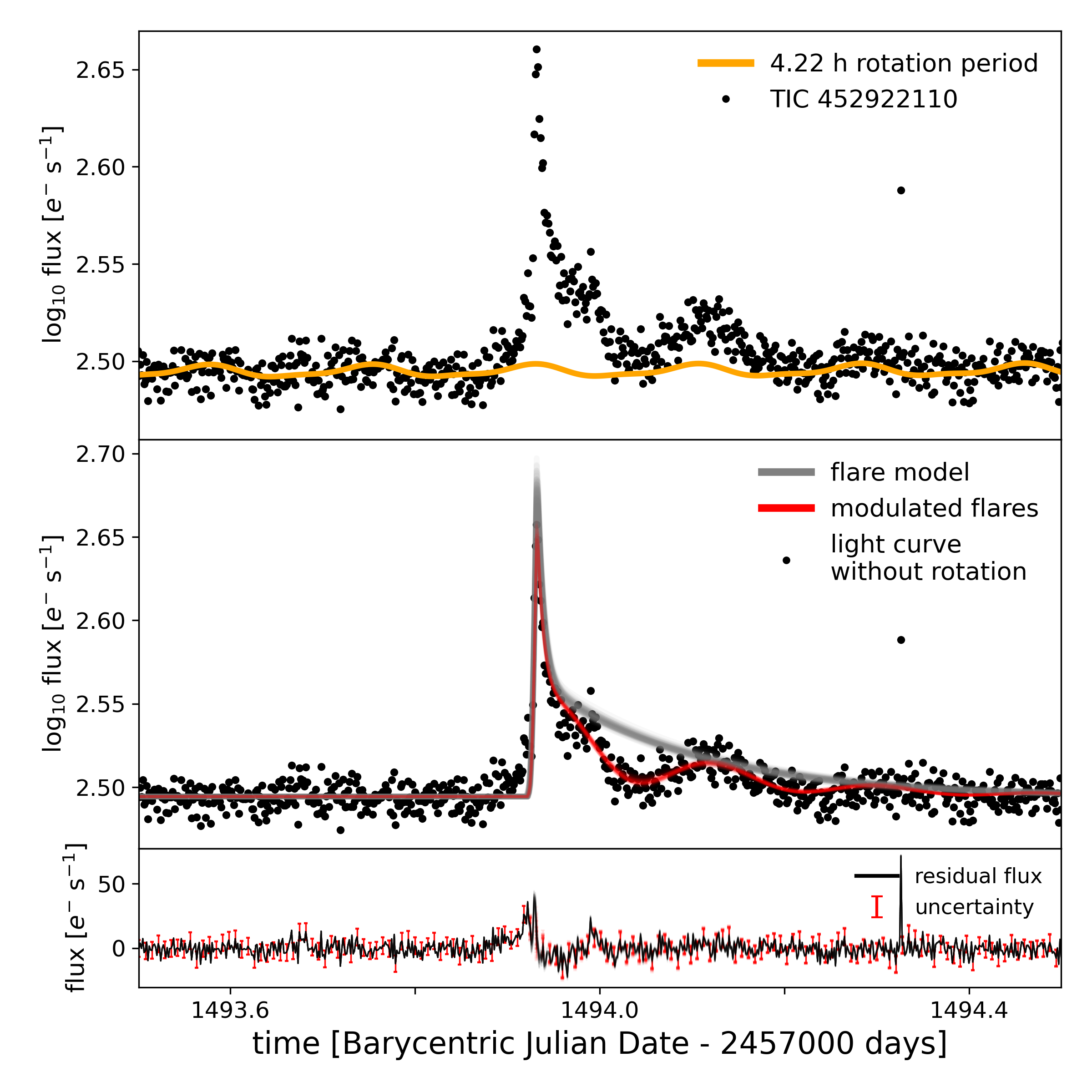}}
\caption{TIC 452: Single flare fit with original uncertainties. The fit yields a flare latitude of $63^\circ$, consistent with the fit in Fig.~\ref{fig:Fig.1} B that adopts increased uncertainty on the flare peak.}
\label{fig:Fig.SM9}
\end{figure}

\begin{figure}
\centerline{\includegraphics[width=0.9 \hsize]{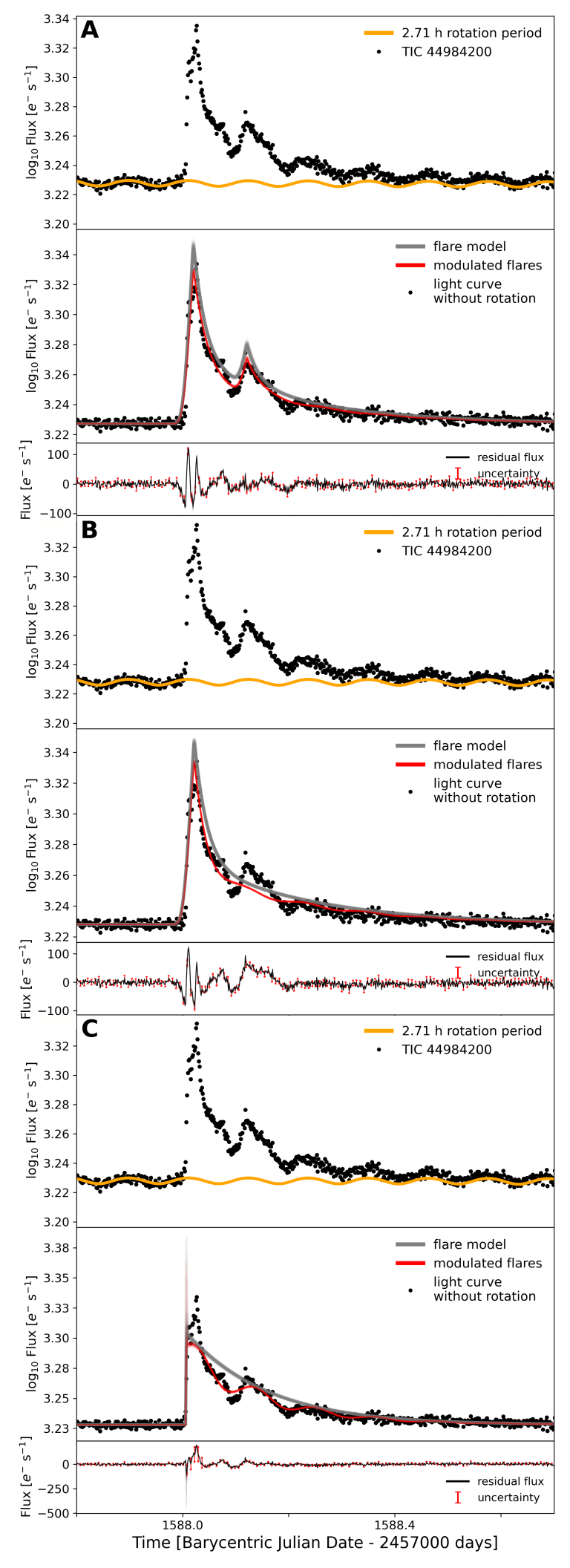}}
\caption{Alternative and ruled out fits to the long-duration flare on TIC 449. \textbf{A}: two-flare fit with original uncertainties. The rotational modulation in the decay phase is underestimated, leading to an overestimated flare latitude of $\sim 86^\circ$. \textbf{B}: single flare fit with original uncertainties. This solution did not capture the secondary peak, or the degree of modulation in the decay phase. Solutions A and B were ruled out. \textbf{C}: single flare fit with increased uncertainties. Relaxing the constraints posed by the impulsive flare peak led to an improved fit of the secondary peak and the decay phase over the solution in B. The fit suggests $\theta_f\approx 70^\circ$, $\sim2^\circ$ lower than our preferred estimate~(Fig.~\ref{fig:Fig.1} C).}
\label{fig:Fig.SM6}
\end{figure}

\begin{figure}
\centerline{\includegraphics[width= \hsize]{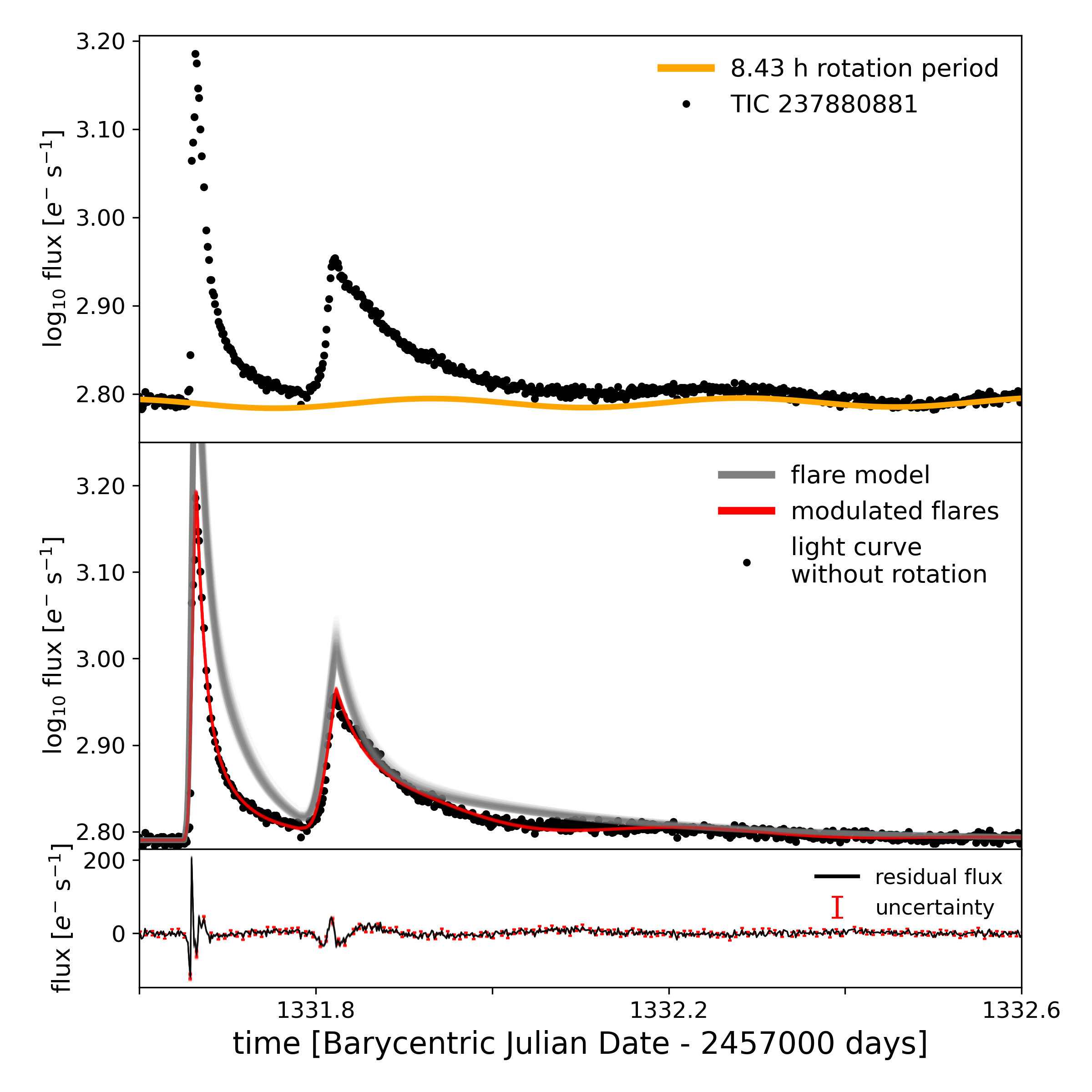}}
\caption{TIC 237: two-flare fit with original uncertainties. The decay phase of the second long duration flare is not well represented by the best fit solution, suggesting a rotational modulation that is too strong. This leads to an underestimated flare latitude of $48^\circ$. This solution was ruled out.}
\label{fig:Fig.SM5}
\end{figure}

In our analysis we used different model setups to fit the light curves. The results we achieved using setups that we ruled out or disfavoured are summarized in Table~\ref{tab:2}, and illustrated in Figs.~\ref{fig:Fig.SM9} \ref{fig:Fig.SM6}, and \ref{fig:Fig.SM5} for TIC 452, TIC 449, and TIC 237, respectively.
\begin{table*}
\caption{Best-fitting results for stars with rotationally modulated superflares that were ruled out or disfavoured. Description of the table rows is the same as in Table~\ref{tab:1}. "h. u." stands for \textbf{h}igh \textbf{u}ncertainties around the peak in the single flare fit to TIC 449. In the remaining fits, we kept the original uncertainties.}
\begin{tabular}{l|cccccc}
\hline\hline
{} &                                            TIC 452 &                                            TIC 449 &                                   TIC 449 (h. u.) &                        TIC 449 (2-flare) &                        TIC 237 (2-flare) \\
\hline
$i$ (deg)                 &                  $49.2\left(^{+3.3}_{-3.3}\right)$ &                  $33.1\left(^{+0.8}_{-0.8}\right)$ &                  $33.1\left(^{+0.8}_{-0.8}\right)$ &        $33.1\left(^{+0.8}_{-0.8}\right)$ &        $25.0\left(^{+4.3}_{-3.3}\right)$ \\
$\log_{10} E_{f,1}$ (erg) &            $33.546\left(^{+0.024}_{-0.021}\right)$ &            $33.364\left(^{+0.005}_{-0.005}\right)$ &            $33.342\left(^{+0.005}_{-0.005}\right)$ &  $33.342\left(^{+0.006}_{-0.006}\right)$ &  $34.765\left(^{+0.034}_{-0.024}\right)$ \\
$A_1$                     &             $0.574\left(^{+0.033}_{-0.027}\right)$ &             $0.324\left(^{+0.004}_{-0.004}\right)$ &             $0.629\left(^{+0.009}_{-0.009}\right)$ &   $0.319\left(^{+0.005}_{-0.004}\right)$ &   $3.507\left(^{+0.303}_{-0.180}\right)$ \\
FWHM$_{i,1}$ (min)        &                  $14.3\left(^{+1.1}_{-0.9}\right)$ &                  $50.3\left(^{+0.8}_{-0.8}\right)$ &                   $0.6\left(^{+0.3}_{-0.3}\right)$ &        $48.4\left(^{+0.8}_{-0.8}\right)$ &        $17.3\left(^{+0.1}_{-0.1}\right)$ \\
FWHM$_{g,1}$ (min)        &                  $64.2\left(^{+3.1}_{-3.0}\right)$ &                  $86.4\left(^{+1.2}_{-1.2}\right)$ &                  $57.6\left(^{+0.8}_{-0.8}\right)$ &        $80.6\left(^{+1.4}_{-1.4}\right)$ &        $16.6\left(^{+0.3}_{-0.3}\right)$ \\
$\log_{10} E_{f,2}$ (erg) &                                                ... &                                                ... &                                                ... &  $32.277\left(^{+0.023}_{-0.022}\right)$ &  $34.712\left(^{+0.030}_{-0.017}\right)$ \\
$A_2$                     &                                                ... &                                                ... &                                                ... &   $0.069\left(^{+0.003}_{-0.003}\right)$ &   $0.660\left(^{+0.047}_{-0.025}\right)$ \\
FWHM$_{i,2}$ (min)        &                                                ... &                                                ... &                                                ... &        $42.9\left(^{+3.3}_{-3.1}\right)$ &        $58.0\left(^{+1.2}_{-1.2}\right)$ \\
FWHM$_{g,2}$ (min)        &                                                ... &                                                ... &                                                ... &        $16.1\left(^{+2.5}_{-2.2}\right)$ &        $96.1\left(^{+1.6}_{-1.6}\right)$ \\
$\theta_f$ (deg)          &                  $63.4\left(^{+3.2}_{-3.5}\right)$ &                  $83.2\left(^{+0.8}_{-0.8}\right)$ &                  $70.0\left(^{+1.0}_{-1.0}\right)$ &        $85.8\left(^{+1.0}_{-1.0}\right)$ &        $48.3\left(^{+5.5}_{-4.8}\right)$ \\
\hline

\end{tabular}

\label{tab:2}
\end{table*}

\section{Limb darkening}
\label{appendix:limb}
The geometric flare modulation model does not account for limb darkening effects that can affect the shape of the modulated flare light curve. The effect increases with flare region size: The larger the fraction of the stellar disk that covered by the flaring region the more do differences in observed stellar flux across the disk take effect.

\begin{figure}
\centerline{\includegraphics[width= \hsize]{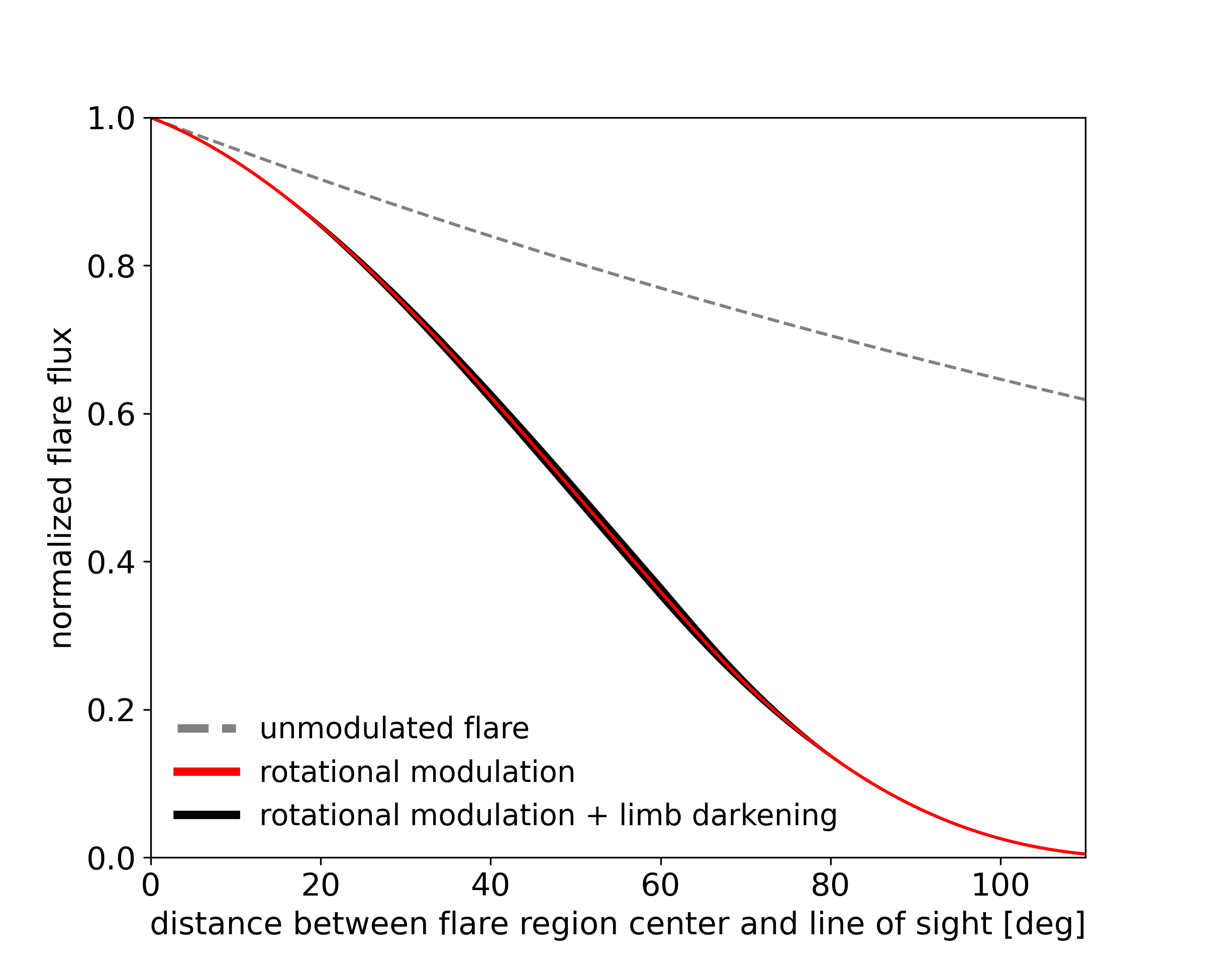}}
\caption{Linear limb darkening applied to an exponential decay of a large (angular radius of $30^\circ$) flaring region that moves across the equator from the line of sight to and over the limb. Grey dashed line: unmodulated exponential flare decay with an e-folding time scale equal to one full stellar rotation. Red line: geometric foreshortening applied to the unmodulated flare. Black area: Geometric foreshortening and linear limb darkening applied to unmodulated flare with limb darkening coefficients between $0.5$ and $1.5$.}
\label{fig:Fig.limbdarkening}
\end{figure}

In Fig.~\ref{fig:Fig.limbdarkening}, we illustrate a scenario that represents the flare with the largest flaring region in our sample (first flare on TIC 237), which is comparable to some of the largest region sizes derived for flares on fully convective dwarfs in the literature~\citep{schmidt2014b, xin2021}. The effect is negligible compared to the noise in the light curves in our sample.

\section{MCMC model fits}
\label{appendix:mcmc}
\begin{figure*}
\centerline{\includegraphics[width= \hsize]{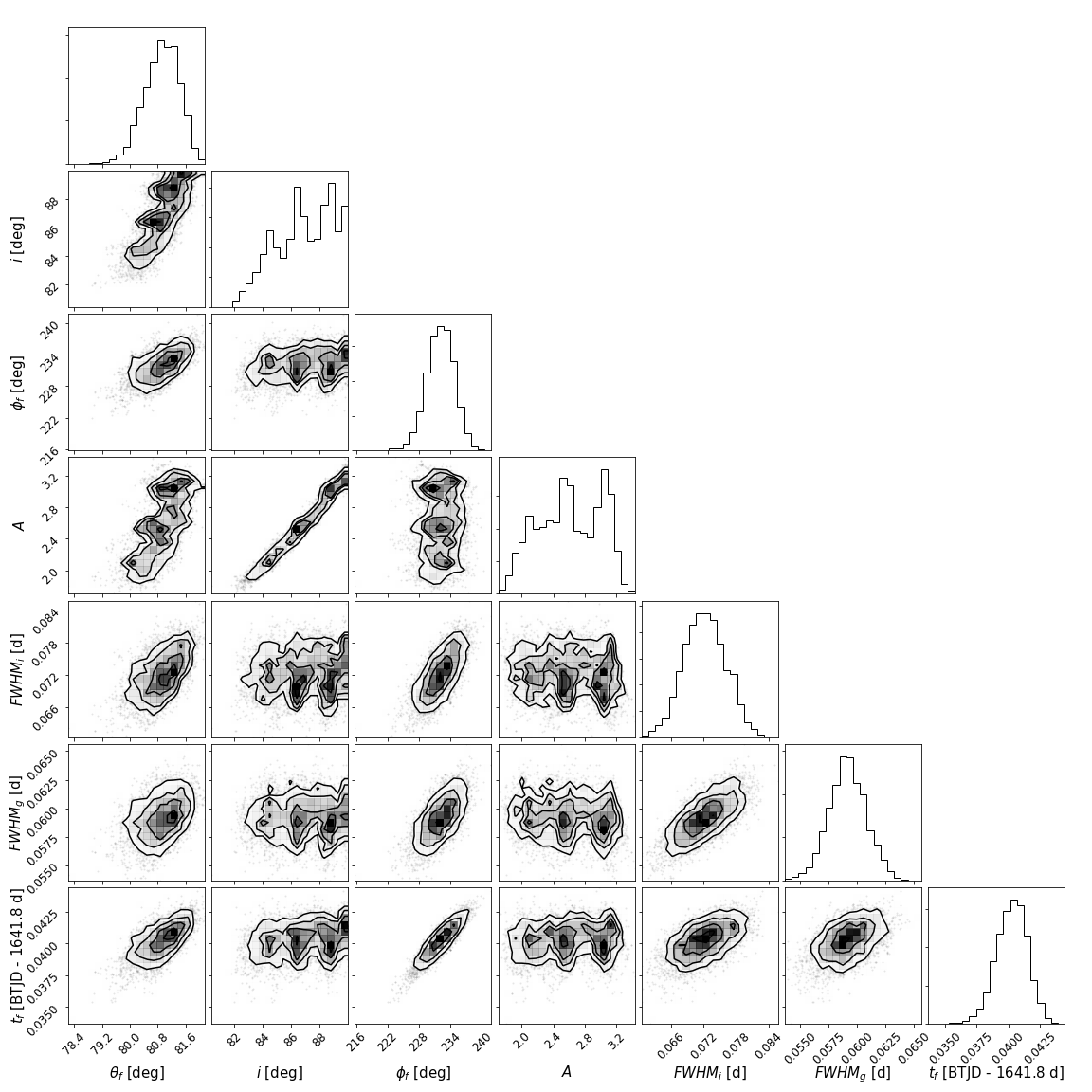}}
\caption{MCMC posterior distributions (density plots) and marginalized distributions (histograms) for the flare fit to the light curve of TIC 277 in Fig.~\ref{fig:Fig.1} A.}
\label{fig:Fig.277mcmc}
\end{figure*}

\begin{figure*}
\centerline{\includegraphics[width= \hsize]{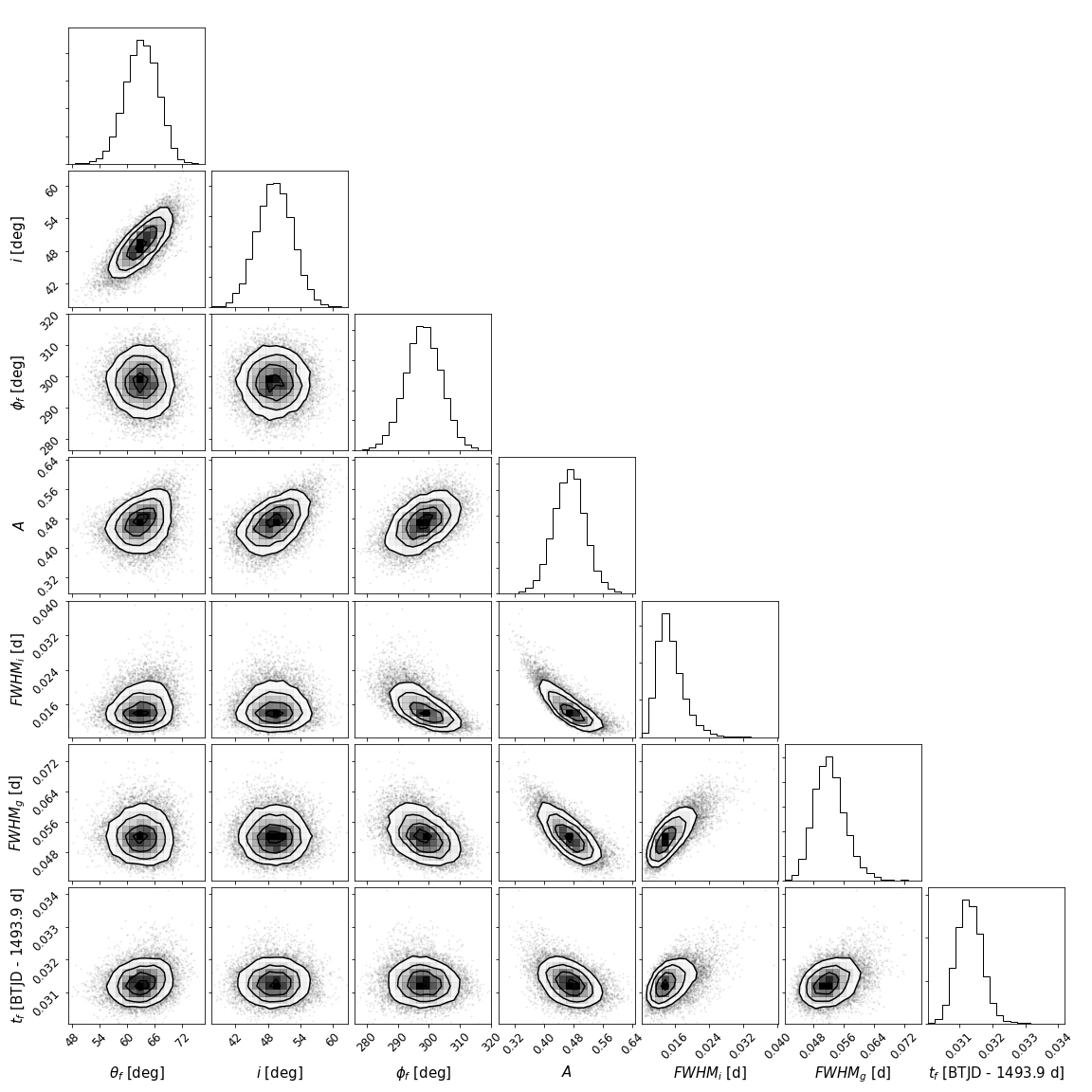}}
\caption{MCMC posterior distributions (density plots) and marginalized distributions (histograms) for the flare fit  to the light curve of TIC 452 in Fig.~\ref{fig:Fig.1} B.}
\label{fig:Fig.452mcmc}
\end{figure*}

\begin{figure*}
\centerline{\includegraphics[width= \hsize]{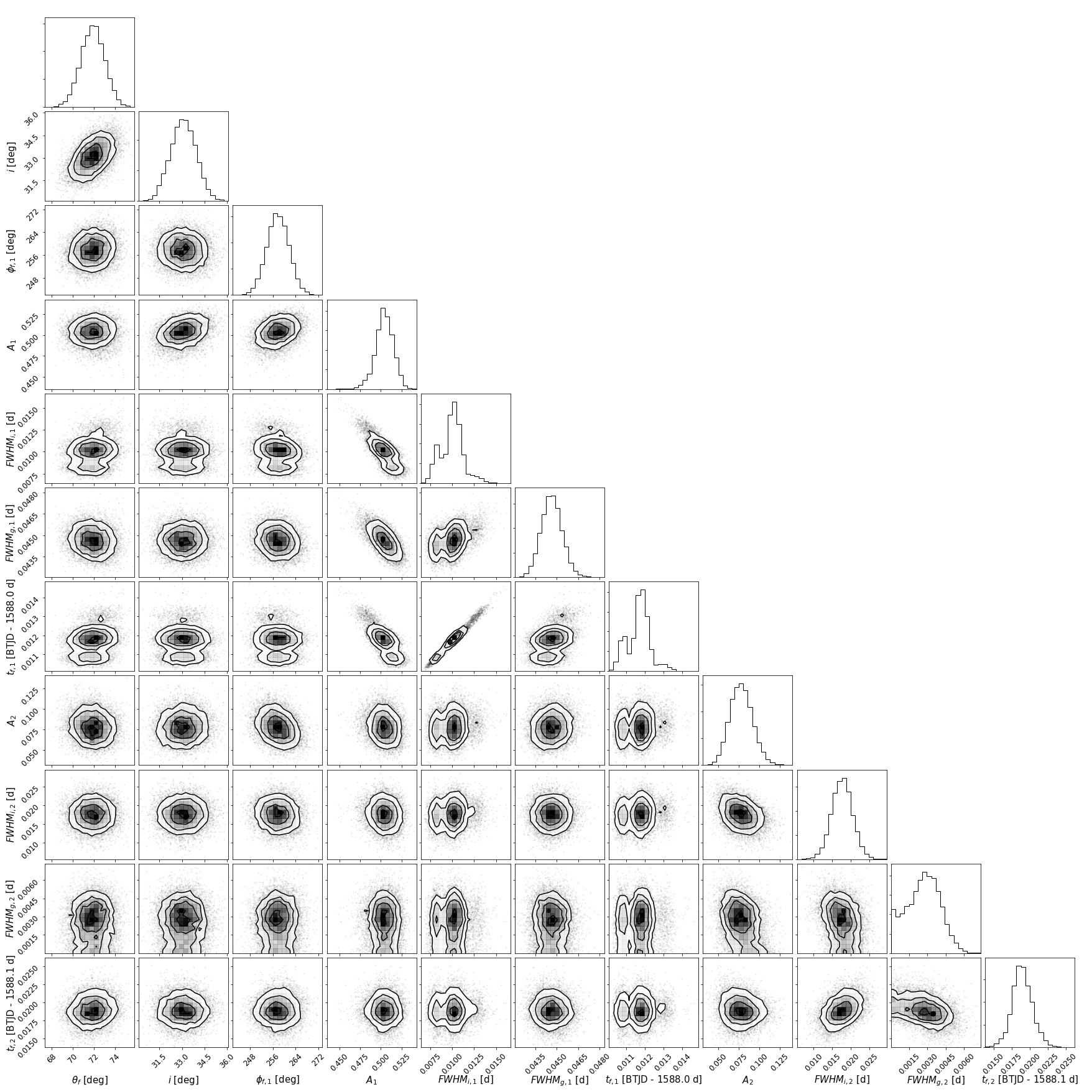}}
\caption{MCMC posterior distributions (density plots) and marginalized distributions (histograms) for the flare fit to the light curve of TIC 449 in Fig.~\ref{fig:Fig.1} C.}
\label{fig:Fig.449mcmc}
\end{figure*}

\begin{figure*}
\centerline{\includegraphics[width= \hsize]{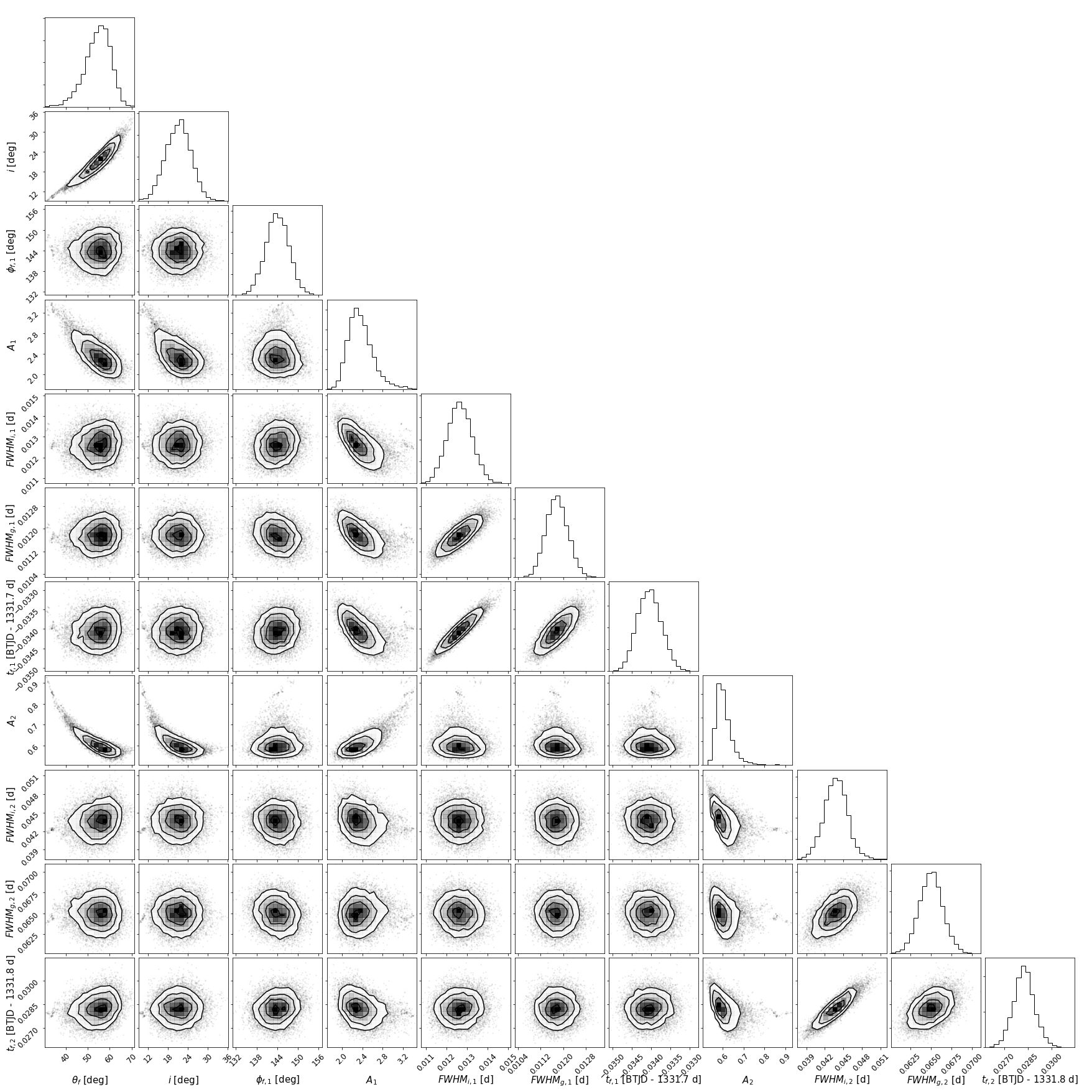}}
\caption{MCMC posterior distributions (density plots) and marginalized distributions (histograms) for the flare fit to the light curve of TIC 237 in Fig.~\ref{fig:Fig.1} D.}
\label{fig:Fig.237mcmc}
\end{figure*}

In Figs.~\ref{fig:Fig.277mcmc}--\ref{fig:Fig.237mcmc}, we show the posterior distributions for the MCMC fits with the best-fit results quoted in Table~\ref{tab:1}. The upper left density plot in each figure shows the partial degeneracy between $\theta_f$ and $i$, which is best seen in TIC 277~(Fig.~\ref{fig:Fig.277mcmc}), where the empirical prior for $i$ was non-Gaussian. In the distributions for TIC 277, TIC 449, and TIC 237, there are some correlations worth noting:

The flare parameters $A_1$, FWHM$_{i,1}$, and FWHM$_{g,1}$ of the first event in TIC 449~(Fig.~\ref{fig:Fig.449mcmc}) are sensitive to the uncertainty on the flare peak time $t_{f,1}$, but the flare latitude is not affected. The tail in the distribution of FWHM$_{g,2}$ suggests that the secondary flare could be fit with a single exponential in the decay phase.

In TIC 277, inclination and flare amplitude are correlated. Higher inclination implies stronger geometrical foreshortening. The distribution suggests that there are three particularly favorable solutions, depending on $A$, which seems counterintuitive at first. But since higher $A$ implies a larger flaring region, we can interpret the solutions as combinations of different flare region sizes viewed at certain inclinations near the pole that result in comparable light curve morphologies. 

Finally, in TIC 237, $\theta_f$ has a tail towards lower latitudes that correlates with higher flare amplitudes $A_1$ and $A_2$. However, as we can see in Fig.~\ref{fig:Fig.237latitudes}, rotational modulation in the decay phase of the event would become more and more pronounced with decreasing latitude, thereby breaking the partial degeneracy in $A_1$, $A_2$ and $\theta_f$.

\bsp	
\label{lastpage}
\end{document}